\documentclass[journal=jacsat,manuscript=article]{achemso}
\usepackage{graphicx} 
\usepackage{amsmath}
\usepackage{comment}
\usepackage{xcolor}
\usepackage{graphicx}
\usepackage{amssymb}
\bibstyle{achemso}
\usepackage{soul}

\newcommand \vect[1]{\textbf{#1}}

\definecolor{ykred}{rgb}{1.0, 0.0, 0.0}
\definecolor{jxc}{rgb}{0.0,0.0,1.0}
\definecolor{slgreen}{rgb}{0.5, 0.8, 0.8}

\definecolor{rzblue}{rgb}{0.5, 0.5, 1.0}
\definecolor{blue}{rgb}{0, 0.0, 1} 
\newcommand{\yk}[1]{#1}

\newcommand{\rz}[1]{#1}
\usepackage{braket}
\usepackage{subcaption} 
\title{All-electron Dynamical Bethe-Salpeter Equation for Extended Systems with Atom-centered Orbital Basis}
\author{Ruiyi Zhou}
\affiliation{Department of Chemistry, University of North Carolina at Chapel Hill, Chapel Hill, North Carolina 27599, USA}
\alsoaffiliation{These authors contributed equally to this work}
\author{Songrui Liu}
\affiliation{Department of Chemistry, University of North Carolina at Chapel Hill, Chapel Hill, North Carolina 27599, USA}
\alsoaffiliation{These authors contributed equally to this work}

\author{Jianhang Xu}
\affiliation{Department of Chemistry, University of North Carolina at Chapel Hill, Chapel Hill, North Carolina 27599, USA}

\author{Yi Yao}
\affiliation{Department of Chemistry, University of North Carolina at Chapel Hill, Chapel Hill, North Carolina 27599, USA}
\alsoaffiliation{Thomas Lord Department of Mechanical Engineering and Materials Science, Duke University, Durham, North Carolina 27708, USA}

\author{Yosuke Kanai}
\email{ykanai@unc.edu}
\affiliation{Department of Chemistry, University of North Carolina at Chapel Hill, Chapel Hill, North Carolina 27599, USA}
\alsoaffiliation{Department of Physics and Astronomy, University of North Carolina at Chapel Hill, Chapel Hill, North Carolina 27599, USA}
\begin{document}

\maketitle

\begin{abstract}
Solving Bethe-Salpeter equation (BSE) for the two-particle Green’s function is the most widely used approach
for taking into account the particle-hole (exciton) interaction in electronic excitation in the context
of the many-body theory based on Green's function. 
In BSE calculations, the static approximation to the screened Coulomb interaction kernel is commonly employed. 
However, when the excitonic character is significant as typically indicated by a large exciton binding energy, 
dynamical screening effects become non-negligible, rendering the static approximation questionable. 
Because of the large computational cost due to the dense Brillouin zone integration necessary for convergence, 
solving the dynamical BSE for extended systems remains a significant challenge, especially when combined with $GW$ calculation for the calculation of quasi-particle energies. 
In this work, we formulate the  plane-wave based effective dielectric function method [Zhang, et al., Phys. Rev. B 107, 235205 (2023)] for the dynamical BSE calculation using atom-centered orbitals as basis functions.
We implement this approach in our recently developed 
all-electron numerical atom-centered orbital (NAO) implementation of BSE@$GW$ [Zhou, et al. J. Chem. Theory Comput. 21, 291 (2025)] for extended systems.
We validate our all-electron NAO-based  implementation of the dynamical BSE method, and then discuss its realistic application to a molecular crystal of naphthalene by performing the dynamical BSE@$G_0W_0$ calculation.
\end{abstract}
\section{Introduction}
In the last decade, the many-body perturbation theory based on Green's function formalism has found its way into chemistry community from condensed matter physics. 
A number of groups have shown that the so-called $GW$ and Bethe-Salpeter equation (BSE) methods can be made quite promising for studying excited state properties of isolated molecules \cite{van2013gw,leng2016gw,golze2019gw, blase2020bethe,holzer2021gw,holzer2025guide,loos2020quest}.
Solving the particle-hole two-particle Green's function, the BSE method \cite{strinati1984effects,albrecht1998ab,rohlfing1998electron,onida2002electronic} has become increasingly popular for calculating neutral excitations of molecules in recent years \cite{blase2020bethe}, adding to a history of successes for condensed matter systems\cite{marsili2017large,vorwerk2019bethe,deslippe2012berkeleygw,sharifzadeh2018many,liu2025many}. It is now widely recognized as a promising alternative\cite{tiago2005first,faber2014excited,hirose2015all,jacquemin2017bethe, blase2018bethe,tolle2021subsystem,franzke2022nmr,graml2025optical} to density functional theory (DFT)-based approaches such as linear-response time-dependent density functional theory (LR-TDDFT)\cite{casida1995time,ullrich2011time} and traditional wavefunction-based methods like the equation-of-motion coupled cluster (EOM-CC) \cite{bartlett2012coupled,krylov2008equation}.
In the last several years, we have formulated and demonstrated all-electron BSE@$GW$ methods using numeric atom-centered orbital (NAO) basis\cite{blum2009ab} for molecules\cite{liu2020all,bhattacharya2024bse}, core electron excitation\cite{yao2022all}, and extended systems\cite{zhou2024all, bhattacharya2026proton}.

Solving the BSE for the two-particle Green's function is the most widely used approach to take into account the electron-hole interaction for electronic excitation in the context of many-body theory. 
Time-dependence in the screened Coulomb interaction is usually assumed to be instantaneous, and this amounts to neglecting the frequency-dependence of the screened Coulomb interaction kernel when solving the BSE eigenvalue problem, similar in spirit to adopting the adiabatic approximation for the exchange-correlation kernel in LR-TDDFT.
This static approximation physically corresponds to assuming the correlations between particles in the pair excitation process are instantaneous while the correlation time is in fact short but finite, related to the period of plasma oscillations\cite{bechstedt_many_2015}.
This static approximation is generally found to work well \rz{when the exciton binding energy is much smaller than the plasma frequency}. 
However, there are situations
where this standard static BSE approximation may not be appropriate.
Dynamical screening effects have been found to influence the excitation energies and spectral shapes appreciably, for core excitations \cite{strinati1984effects}, in metallic systems \cite{marini2003dynamical}, doped two-dimensional systems \cite{gao2016dynamical} and organic systems \cite{bintrim2022full, zhang2023effect}. 
As discussed in the following, this is of particular importance in the context of chemistry application because the dynamical screening tends to be more important for systems with strong excitonic characters as often found for molecules\cite{wen2026dynamically}, nano-scale/low-dimensional systems,\cite{qiu2016screening,gao2016dynamical,zhu2015exciton,ugeda2014giant}  and in organic solids\cite{dadsetani2015ab,liu2020pyrene,wang2016effect,hummer2005oligoacene,bhattacharya2026proton}.
Among a few different approaches to incorporate dynamical screening effects in BSE calculation, the effective dielectric function method\cite{zhang2023effect} by \citeauthor{zhang2023effect} is particularly attractive from the viewpoint of computational cost.
This is a particularly important consideration because BSE@$GW$ method tends to be quite intensive computationally for extended systems due to the dense sampling of the Brillouin zone required for convergence\cite{ deslippe2012berkeleygw,kammerlander2012speeding,alvertis2023importance,zhou2024all}.
Building on our recent efforts on developing all-electron numeric atom-centered orbital (NAO) basis set approach for BSE@$GW$\cite{zhou2024all,liu2020all,yao2022all}, we develop this 
effective dielectric function approach for dynamical BSE, which was originally formulated using plane-waves by \citeauthor{zhang2023effect}, for the all-electron NAO-based BSE@$GW$ method.

\section{Theoretical Methods}
\subsection{Dynamical Bethe-Salpeter Equation and Shindo's Approximation}
Let us concisely review theoretical foundation since the dynamical, as opposed to static, formulation of Bethe-Salpeter equations (BSE) is 
not widely discussed in literature.
BSE for the macroscopic polarization function $P^M$ is given by
\begin{equation}
    P^M(11',22') = L^0(11',22') + \int \mathrm{d}3 \int \mathrm{d}4 \int \mathrm{d}5 \int \mathrm{d}6\ L^0(11',43) \Xi^M(34,56) P^M(56,22')\label{eq:polar_BSE}
\end{equation}
where the composite indices $1\equiv(\mathbf{x}_1,t_1)\equiv(\vect r_1,m_{s1}, t_1)$, etc., denote spin-space coordinates in time. 
Here, the density correlation function $L^0$ for the propagation of two independent particles can be identified as $L^0(11',22')=-i\hbar G(12')G(21')$, and $G(12')$ is the one-particle Green's function.
The kernel $\Xi^M$ is defined as the derivative of the total self-energy $\Sigma$ with respect to the one-particle Green's function, 
\begin{equation}\label{eq:kernel}\Xi^M(34,56)=-\frac{1}{i\hbar}\frac{\delta\Sigma(43)}{\delta G(56)}
\approx \bar v(3,6)\delta(34)\delta(5^+6)-W(4^+3)\delta(36)\delta(45).\end{equation}
where $W$ is the screened Coulomb potential and 
\rz{$\bar v$ denotes the bare Coulomb potential with its $\mathbf{G}=0$ Fourier component removed, $\bar v(\mathbf{q}+\mathbf{G})=v(\mathbf{q}+\mathbf{G})\left(1-\delta_{\mathbf{G},\mathbf{0}}\right)$. Removing this component makes the kernel generate the \textit{macroscopic} polarization function $P^M$ rather than the full polarization function. 
} 
Here, the exchange-correlation part of the kernel is approximated by the $GW$ approximation, and the functional derivative $\delta W/\delta G$ is neglected.
With Fourier transformation, the macroscopic polarization function (Eq. \ref{eq:polar_BSE}) depends on two frequencies $\omega_n$ and $\tilde{\omega}_m$, 
 
\begin{equation}
\label{eq:dble_freq_p}
\begin{aligned}
P^M(\vect{x}_1\vect{x}_1^{\prime},\vect{x}_{2}\vect{x}_{2}^{\prime};\omega_n\tilde{\omega}_m)&=- i\hbar 
\biggr\{G(\vect{x}_{1}\vect{x}_{2}',\omega_n)G(\vect{x}_{2}\vect{x}_{1}',\omega_n-\tilde{\omega}_m)\\&
+\frac{1}{i\hbar\beta}
\sum_{n^{\prime}} \sum_{m_{s3}, m_{s4}} \sum_{m_{s3}', m_{s4}'}
\int \mathrm{d}\vect{x}_{3}\ \mathrm{d}\vect{x}_{4}\ \mathrm{d}\vect{x}_{3}'\ \mathrm{d}\vect{x}_{4}'\ G(\vect{x}_{1}\vect{x}_{3},\omega_n)
\\&\times G(\vect{x}_{4}\vect{x}_{1}',\omega_n-\tilde{\omega}_m)
\Big[-\delta_{m_{s3}m_{s3}'}\delta_{m_{s4}m_{s4}'}W({\vect{x}_{3}}\vect{x}_{4}, \omega_{n}-\omega_{n^{\prime}})\\&+ \delta_{m_{s3}m_{s4}}\delta_{m_{s3}'m_{s4}'}\bar v({\vect{x}_{3}\vect{x}_{4}')}\Big]P^M(\vect{x}_3'\vect{x}_4',\vect{x}_2\vect{x}_{2^{\prime}};\omega_{n^{\prime}}\tilde{\omega}_m)\biggr\}
\end{aligned}\end{equation} 
where $\beta$ is the inverse temperature and $W(\vect{x}\vect{x}',\omega_n)=\int_{0}^{-i\hbar\beta}dt e^{i \omega_n(t-t')}W(\vect{x}\vect{x}',t-t')$.
The frequency $\omega_n$ corresponds to the Fermionic Matsubara frequency from the time difference $t_1-t_1'$ while $\tilde{\omega}_m$ is the Bosonic Matsubara frequency of time difference $t_1- t_2$\cite{bechstedt_many_2015}.
The \textit{single}-frequency polarization needed for the optical excitation (the macroscopic dielectric function) can be, in principle, obtained from Eq. \ref{eq:dble_freq_p} as 
\begin{equation}
\label{eq:single_double_p}
P^M(\vect{x}_1\vect{x}_1^{\prime},\vect{x}_{2}\vect{x}_{2}^{\prime};\tilde{\omega}_m)=\frac{1}{-i\hbar\beta}\sum_n{P^M(\vect{x}_1\vect{x}_1^{\prime},\vect{x}_{2}\vect{x}_{2}^{\prime};\omega_n\tilde{\omega}_m)}.
\end{equation}
In practice, however, evaluating Eq. \ref{eq:single_double_p} is  numerically challenging as discussed in detail by \citeauthor{bechstedt_many_2015}\cite{bechstedt_many_2015}.
The most commonly used approach to avoid this practical difficulty is to simply neglect the frequency dependence of the screened Coulomb interaction
(i.e. $W({\vect{x}_{1}}\vect{x}_{2}, \omega_{n}-\omega_{n^{\prime}})= W({\vect{x}_{1}}\vect{x}_{2}, 0)$) in Eq.~\ref{eq:dble_freq_p}.
With this approximation, the usual \textit{static} BSE cast as a generalized eigenvalue equation is obtained as typically found in literature. 

Let us now discuss how Shindo's approximation\cite{shindo_effective_1970} enables us to circumvent the above-mentioned difficulty in calculating the polarization function as this approximation forms the foundation of the work\cite{zhang2023effect} by \citeauthor{zhang2023effect}.
For convenience, we express the one-particle Green's function 
by substituting the spin-space representation with \rz{the orbitals $\psi_\lambda$ and an explicit spin index $m_s$},  restricting ourselves to the collinear spins and singlet excitation, 
\begin{align}
G^{m_sm_s'}_{\lambda\lambda'}(\omega_n) &= \iint \rm d \vect x\ \rm d \vect x' \ \psi_\lambda(\vect{x})G(\vect{x}\vect{x}',\omega_n)\psi^{*}_{\lambda'}(\vect{x}') 
    = \frac{\delta_{m_sm_s'}\delta_{\lambda\lambda'} }{\hbar \omega_n - \varepsilon_{\lambda}^{\mathrm{QP}}}    
    =G_{\lambda}(\omega_n)\delta_{m_sm_s'}\delta_{\lambda\lambda'},
\end{align}
where we used
$G_{\lambda}(\omega) = (\hbar \omega - \varepsilon_{\lambda}^{\mathrm{QP}})^{-1}$, and $\varepsilon_{\lambda}^{\mathrm{QP}}$ is the quasiparticle energy, which can be obtained by solving the Dyson equation.
In the so-called Shindo's approximation\cite{shindo_effective_1970, schafer1986approach}, 
valid in the limit of weak dynamical screening influence (see Ref. \citenum{bechstedt_many_2015} for derivation), 
the two-frequency polarization function is approximated as

\begin{equation}
\label{eq:shindo}
P^M(\lambda_1\lambda_1^{\prime},\lambda_2\lambda_2^{\prime};\omega_n\tilde{\omega}_m)\approx i\hbar\times\frac{G_{\lambda_1}(\omega_n)-G_{\lambda_1^{\prime}}(\omega_n-\tilde{\omega}_m)}{\tilde{N}_{\lambda_{1}\lambda_{1}^{\prime}}(\tilde{\omega}_{m})}P^M(\lambda_1\lambda_1^{\prime},\lambda_2\lambda_2^{\prime};\tilde{\omega}_m).
\end{equation}
The normalization factor, $\tilde{N}_{\lambda_{1}\lambda_{1}^{\prime}}(\tilde{\omega}_{m}) =  -{1}/\beta\sum_{n}[G_{\lambda_1}(\omega_{n})-G_{\lambda_1^{\prime}}(\omega_{n}-\tilde{\omega}_m)]$, can be  given in terms of the Fermi-Dirac distribution at thermal equilibrium as $\tilde{N}_{\lambda_{1}\lambda_{1}^{\prime}}(\tilde{\omega}_{m}) =f(\varepsilon^{QP}_{\lambda_{1}^{\prime}})-f(\varepsilon^{QP}_{\lambda_{1}})$
without explicit frequency dependence.
The key point is that
the Shindo’s approximation uncouples the frequency dependencies in the two-frequency dependent polarization function by using the Green’s functions for single-particle states. 
With the Shindo's approximation, Eq. \ref{eq:dble_freq_p}  simplifies to a single-frequency dependent polarization BSE, 
\begin{equation}\begin{aligned}    
&\left[\varepsilon_{\lambda_{1}}^{\mathrm{QP}}-\varepsilon_{\lambda_{1}^{\prime}}^{\mathrm{QP}}-\hbar\tilde{\omega}_{m}\right]P^{M}(\lambda_{1}\lambda_{1}^{\prime},\lambda_{2}\lambda_{2}^{\prime};\tilde{\omega}_{m})-\sum_{\lambda_{3}\lambda_{4}}\left[\tilde{N}_{\lambda_{1}\lambda_{1}^{\prime}}(\tilde{\omega}_{m})\tilde{W}_{\lambda_{1}^{\prime}\lambda_{4}}^{\lambda_{1}\lambda_{3}}(\tilde{\omega}_{m})-2\bar{{v}}_{\lambda_{3}\lambda_{4}}^{\lambda_{1}\lambda_{1}^{\prime}}\right]\\&\times P^{M}(\lambda_{3}\lambda_{4},\lambda_{2}\lambda_{2}^{\prime};\tilde{\omega}_{m})=-\delta_{\lambda_{1}\lambda_{2}^{\prime}}\delta_{\lambda_{2}\lambda_{1}^{\prime}}\tilde{N}_{\lambda_{1}\lambda_{1}^{\prime}}(\tilde{\omega}_{m}).
\end{aligned}
\label{eq:single-frequency-BSE}
\end{equation}

\rz{Here, $\lambda$ labels an orbital rather than a spin orbital,
and thus the spin summation for the singlet channel has already been carried out at this point, yielding the factor of 2 in front of $\bar v$ in Eq. \ref{eq:single-frequency-BSE}. 
The kernel matrix element $\bar{v}_{\lambda_{3}\lambda_{4}}^{\lambda_{1}\lambda_{1}^{\prime}}$ involves the same operator $\bar v$ that appears in Eq.~\ref{eq:kernel}, evaluated between two such electron-hole transition pairs, $\bar{v}_{\lambda_{3}\lambda_{4}}^{\lambda_{1}\lambda_{1}^{\prime}}=\iint {\rm d}\vect r\ {\rm d}\vect r^{\prime}\ \psi^{*}_{\lambda_{1}}(\vect r)\psi_{\lambda_{1}^{\prime}}(\vect r)\,\bar v(\vect r,\vect r^{\prime})\,\psi_{\lambda_{3}}(\vect r^{\prime})\psi^{*}_{\lambda_{4}}(\vect r^{\prime})$. 
The dynamically screened Coulomb interaction $\tilde{W}_{\lambda_{1}^{\prime}\lambda_{4}}^{\lambda_{1}\lambda_{3}}$ is built in the same way, except that $\lambda_{1}$ is now paired with $\lambda_{3}$ instead of with $\lambda_{1}^{\prime}$. 
This difference in the pairing is the usual distinction between the exchange and the direct term of the BSE kernel\cite{bechstedt_many_2015,onida2002electronic}. }
The  frequency-dependent dynamically screened Coulomb potential, $\tilde{W}_{\lambda_{2}\lambda_{4}}^{\lambda_{1}\lambda_{3}}(\tilde{\omega}_{m})$, is given by
\begin{equation}
\begin{aligned}
\label{eq:w_matsubara1}
\tilde{W}_{\lambda_2\lambda_4}^{\lambda_1\lambda_3}(\tilde \omega_m) 
&= \frac{1}{\beta^2} \sum_{n,n^{\prime}} 
  \frac{[G_{\lambda_1}(\omega_n) - G_{\lambda_2}(\omega_n - \tilde \omega_m)] \times [G_{\lambda_3}(\omega_{n^{\prime}}) - G_{\lambda_4}(\omega_{n^{\prime}} - \tilde \omega_m)]}{
    \tilde{N}_{\lambda_1\lambda_2}(\tilde \omega_m)
    \tilde{N}_{\lambda_3\lambda_4}(\tilde \omega_m)
  }\\&\times\iint  {\rm d} \vect r\ {\rm d} \vect r' \ \psi^{*}_{\lambda_1}(\vect{r})\psi_{\lambda_3}(\vect{r})\psi_{\lambda_2}(\vect{r}')\psi^{*}_{\lambda_4}(\vect{r}')W(\vect{r}\vect{r}',\omega_n-\omega_{n'}).
  \end{aligned}
\end{equation}
This expression can be  simplified by expressing the screened Coulomb term as 
\begin{equation}
\begin{aligned}
\label{eq:w_matsubara2}
W(\vect{r}\vect{r}',\omega_n-\omega_{n'})
&
=v(\mathbf{r},\mathbf{r}^{\prime})
\left\{
\delta(\mathbf{r},\mathbf{r}^{\prime})+\int\limits_{-\infty}^{\infty}\frac{d\omega^{\prime}}{\pi}
\frac{\mathrm{Im}\epsilon^{-1}(\mathbf{r}\mathbf{r}^{\prime},\omega^{\prime})}{\omega^{\prime}-\tilde \omega_m}\right\} 
\end{aligned}
\end{equation}
where $\epsilon^{-1}(\mathbf{r}\mathbf{r}^{\prime},\omega^{\prime})$ is the inverse dielectric function.
To formulate this equation in a more practical form for first-principles calculations of extended periodic systems, 
we write it in terms of the Bloch states ($\lambda \rightarrow v\vect k, c\vect k$, which denotes the valence band and conduction band, respectively), considering translationally invariant systems, and the interband normalization factors are $\tilde{N}_{cv\mathbf{k}}(\tilde{\omega}_m)=-\tilde{N}_{vc\mathbf{k}}(\tilde{\omega}_m)=1$.
With the analytic continuation $\tilde{\omega}_m \rightarrow \omega$, the dynamically screened Coulomb interaction is \begin{equation}
\begin{aligned}
\label{eq:w_1}
\braket{ v\mathbf{k} v^{\prime}\mathbf{k}^{\prime}|\hat{W}(\omega)|c\mathbf{k}c^{\prime}\mathbf{k}^{\prime}} 
&=\iint d \mathbf{r}d \mathbf{r^\prime }
v(\mathbf{r},\mathbf{r}^{\prime})
\psi^{*}_{c\mathbf{k}}(\vect{r})\psi_{c'\mathbf{k}'}(\vect{r})
\psi_{v\mathbf{k}}(\vect{r}')\psi^*_{v'\mathbf{k}'}(\vect{r}') \\
&\times \Bigg\{
\delta(\mathbf{r},\mathbf{r}^{\prime})+\int_{0}^{\infty}\frac{d\hbar\omega^{\prime}}{\pi}
\,\mathrm{Im}\,\epsilon^{-1}
(\mathbf{r}\mathbf{r^{\prime}},\omega^{\prime})\\
&\qquad\times
\left[
\frac{1}{
\hbar\omega^{\prime}
+\varepsilon_{c\mathbf{k}}^{\mathrm{QP}}
-\varepsilon_{v^{\prime}\mathbf{k}^{\prime}}^{\mathrm{QP}}
-\hbar \omega}
+\frac{1}{
\hbar\omega^{\prime}
+\varepsilon_{c^{\prime}\mathbf{k}^{\prime}}^{\mathrm{QP}}
-\varepsilon_{v\mathbf{k}}^{\mathrm{QP}}
-\hbar \omega}
\right]
\Bigg\},
\end{aligned}
\end{equation}
%
where indices $c,c^{\prime}$ represent the conduction band (CB) states, $v,v^{\prime}$ represent the valence band (VB) states, and ${\mathbf k}, {\mathbf k}^{\prime}$ represent the $k$-point in the first Brillouin zone. 
As noted by \citeauthor{bechstedt_many_2015}\cite{bechstedt_many_2015}, \rz{the dynamical effects on the exciton become more pronounced as $|\hbar \omega-(\varepsilon_{c}^{\mathrm{QP}}-\varepsilon_{v}^{\mathrm{QP}})|$, which represents the exciton binding energy, becomes larger relative to the plasma frequency, and the static BSE becomes correspondingly less reliable.}
Such situations are generally suspected for molecular systems and core excitations even in extended solids due to their larger exciton binding energy, and these are precisely the systems for which our all-electron NAO-based BSE@$GW$ method\cite{liu2020all,yao2022all, zhou2024all} could be particularly suitable.
Within the standard Tamm-Dancoff approximation (TDA)\cite{hirata1999time}, 
the exciton Hamiltonian matrix for the frequency-dependent BSE (Eq. \ref{eq:single-frequency-BSE}) can be written as
\begin{equation}
\label{eq:bse_full_freq}
\begin{aligned}
\langle v\mathbf{k}c\mathbf{k}|\vect{A}(\omega)|v^{\prime}\mathbf{k}^{\prime}c^{\prime}\mathbf{k}^{\prime}\rangle&=(\epsilon_{c\mathbf{k}}^{\mathrm{QP}}-\epsilon_{v\mathbf{k}}^{\mathrm{QP}})\delta_{vv^{\prime}}\delta_{cc^{\prime}}\delta_{\mathbf{k}\mathbf{k}^{\prime}}
+2\langle v\mathbf{k}c\mathbf{k}|\hat{V}|v^{\prime}\mathbf{k}^{\prime}c^{\prime}\mathbf{k}^{\prime}\rangle
-\langle v\mathbf{k}v^{\prime}\mathbf{k}^{\prime}|\hat{W}(\omega)|c\mathbf{k}c^{\prime}\mathbf{k}^{\prime}\rangle
\end{aligned}
\end{equation}
where operators $\hat{V}$ and $\hat{W}(\omega)$ represent the bare Coulomb operator and frequency-dependent screened Coulomb operator, respectively.
Instead of the standard eigenvalue problem in the static BSE case, the frequency-dependence requires solving a coupled eigenvalue problem,
\begin{equation}\label{eq:dynamic_BSE}
\begin{aligned}
    A(\omega)X_S(\omega) &= E_S(\omega) X_S(\omega) \\
    E_S(\omega) &= \hbar \omega. 
\end{aligned}
\end{equation}
where $E_S$ and $X_S$ represent the excitation energy and eigenvector of the many-body state $S$. Solving these coupled equations (Eq. \ref{eq:dynamic_BSE}) requires that the energy of the frequency-dependent excitonic state, $E_{S}(\omega)$, is equal to the energy of the absorbed photon, $\hbar\omega$.
However, computational methods to solve such coupled eigenvalue problems are often tedious. 
For example, the exact diagonalization method requires sampling the frequency $\omega$ and solving the eigenvalue problem at each sampled frequency point to find the solution. 
We note that there exists an alternative, formally exact approach where the dynamical BSE is reformulated as a frequency-independent eigenvalue problem in an expanded space of single and double excitations  by \citeauthor{bintrim2022full}\cite{bintrim2022full}.
However, its implementation comes with a high-order scaling for the computational cost, limiting its application to small systems and  currently not suitable for extended systems we focus here.

\vskip 20 pt

\noindent
\subsection{Effective Dielectric Function Approach }
Different approximations for incorporating the dynamical screening effects exist. 
The early work by \citeauthor{rohlfing2000electron} treated the dynamical screening as a first-order perturbation\cite{rohlfing2000electron}, and \citeauthor{loos2020dynamical} implemented and benchmarked this method for molecular systems in recent years\cite{loos2020dynamical}.
A relatively recent work by \citeauthor{gao2016dynamical} employed an effective screening function\cite{gao2016dynamical} for studying dynamical excitonic effect in doped 2D materials, and 
this idea was further developed by \citeauthor{zhang2023effect} in Ref. \citenum{zhang2023effect}.
\rz{Such an idea of using an effective screening for dynamical BSE dates back as far as the work in 1984 by \citeauthor{strinati1984effects} who treated core excitons through an effective dielectric matrix\cite{strinati1984effects}.}
The approach by \citeauthor{zhang2023effect} bypasses the frequency-dependent eigenvalue problem by replacing the energy difference terms $\varepsilon_{c\mathbf{k}}^{\mathrm{QP}}-\varepsilon_{v^{\prime}\mathbf{k}^{\prime}}^{\mathrm{QP}}-\hbar \omega$ and $\varepsilon_{c^{\prime}\mathbf{k}^{\prime}}^{\mathrm{QP}}-\varepsilon_{v\mathbf{k}}^{\mathrm{QP}}-\hbar \omega$ in Eq. \ref{eq:w_1} with the exciton binding energy  $E_b$ of the lowest exciton state
\begin{equation}
\label{eq:eb_st}
    E_b = E^{\rm QP}_{g}-E^{sta.}_{S=0}
\end{equation}
where $E^{\rm QP}_{g}$ is the quasiparticle bandgap and $E^{sta.}_{S=0}$ is the lowest solution of the static BSE calculation with the frequency-independent screened Coulomb potential.
Following Bechstedt\cite{bechstedt_many_2015}, Eq. $\ref{eq:w_1}$ is reduced to a form 
with no explicit dependence on the frequency,
\begin{equation}
\label{eq:w_eff}
\begin{aligned}
\braket{ v\mathbf{k} v^{\prime}\mathbf{k}^{\prime}|\hat{W}^{\rm eff}|c\mathbf{k}c^{\prime}\mathbf{k}^{\prime}} 
&=\frac{1}{\Omega}
\sum_{\mathbf{q}}\sum_{\mathbf{G},\mathbf{G}^{\prime}}
v\left(\sqrt{|\mathbf{q}+\mathbf{G}||\mathbf{q}+\mathbf{G}^{\prime}|}\right)
B_{c\mathbf{k}}^{c^{\prime}\mathbf{k}^{\prime}}(\mathbf{q}+\mathbf{G})B_{v\mathbf{k}}^{v^{\prime}\mathbf{k}^{\prime}*}(\mathbf{q}+\mathbf{G}^{\prime}) \\
&\times \left\{\delta_{\mathbf{GG}^{\prime}}+\int_0^\infty\frac{d\hbar\omega^{\prime}}{\pi}\mathrm{Im}\varepsilon^{-1}(\mathbf{q}+\mathbf{G},\mathbf{q}+\mathbf{G}^{\prime},\omega^{\prime})\frac{2}{\hbar\omega^{\prime}+E_b}\right\}\delta_{\mathbf{q},\mathbf{k}-\mathbf{k}^{\prime}}
\end{aligned}
\end{equation}
where  $v$ is the Coulomb potential in reciprocal space. $\Omega$ is the volume of the unit cell, $\mathbf{q}$ represents the reciprocal wavevector, $\mathbf{G}$ is a reciprocal lattice vector.
The Bloch integrals are given by 
$ B_{c/v \mathbf{k}}^{c^{\prime}/v^{\prime}\mathbf{k}^{\prime}}(\mathbf{q}+\mathbf{G}) = \int \mathrm{d} \vect{r} \ \psi^{*}_{c/v ,\mathbf{k}}(\vect{r})e^{i(\mathbf{q}+\mathbf{G})\cdot \mathbf{r}}\psi_{c^{\prime}/v^\prime ,\mathbf{k}^{\prime}}(\vect{r})$. 
\rz{Note that the exciton binding energy, $E_b$, enters the dynamical calculation as an input taken from a preceding static BSE calculation. 
The subsequent dynamical BSE calculation gives a different exciton binding energy, and thus this procedure can, in principle, be iterated until convergence. 
We examine the sensitivity to the input exciton binding energy value in the Results section, and it is small enough for the present system that a single step suffices, as also noted in Ref.~\citenum{zhang2023effect}.}

To make this approach more computationally tractable, \citeauthor{zhang2023effect} adapted
the plasmon-pole model\cite{hybertsen1986electron} so that the frequency integration ($\omega^{\prime}$) can be performed analytically in their work
\cite{zhang2023effect} with
\begin{equation*}
\begin{aligned}
{\rm Im}\varepsilon^{-1}(\mathbf{q}+\mathbf{G},\mathbf{q}+\mathbf{G^{\prime}},\omega)&=A(\mathbf{q}+\mathbf{G},\mathbf{q}+\mathbf{G^{\prime}})\\ &\times\{\delta[\omega-\tilde{\omega}(\mathbf{q}+\mathbf{G},\mathbf{q}+\mathbf{G^{\prime}})]-\delta[\omega+\tilde{\omega}(\mathbf{q}+\mathbf{G},\mathbf{q}+\mathbf{G^{\prime}})]\},
\end{aligned}
\end{equation*}
\rz{
where $A(\mathbf q+\mathbf G,\mathbf q+\mathbf G')$ is the amplitude of plasma pole and $\tilde{\omega}(\mathbf{q}+\mathbf{G},\mathbf{q}+\mathbf{G^{\prime}})$ is the pole frequency.}
In their work, only the diagonal matrix elements (i.e., $\mathbf{G} = \mathbf{G}^{\prime}$) are retained in
the dielectric function $\epsilon(\mathbf{q} + \mathbf{G}, \mathbf{q} + \mathbf{G}^{\prime}, \omega)$\cite{bechstedt1992efficient}. 
This approximation effectively neglects the local-field effect in the screening, a treatment that has proven both computationally efficient and physically robust for a wide range of semiconductors and related materials~\cite{schleife2009optical, bechstedt1992efficient, schleife2011electronic}.
\rz{Within this approximation, $A(\mathbf q+ \mathbf G)$ and $\tilde{\omega}(\mathbf{q}+\mathbf{G})$
are determined by the Kramers-Kronig relation, the $f$-sum rule, and the static inverse dielectric function. This gives that
$
A(\mathbf{q}+\mathbf{G})=-\frac{\pi}{2}\omega_p[1-\varepsilon^{-1}(\mathbf{q}+\mathbf{G},\omega=0)]^{1/2}, 
$
and 
$\tilde{\omega}(\mathbf{q}+\mathbf{G})=\omega_p[1-\varepsilon^{-1}(\mathbf{q}+\mathbf{G},\omega=0)]^{-1/2}.$} Then, 
Eq. \ref{eq:w_eff} simplifies to
\begin{equation}
\begin{aligned}
\label{eq:w_zhang}
&\braket{ v\mathbf{k} v^{\prime}\mathbf{k}^{\prime}|\hat{W}^{\rm eff}|c\mathbf{k}c^{\prime}\mathbf{k}^{\prime}} 
=\frac{1}{\Omega}\sum_{\mathbf{q}}\sum_{\mathbf{G}}
v(|\mathbf{q}+\mathbf{G}|)
B_{c\mathbf{k}}^{c^{\prime}\mathbf{k}^{\prime}}(\mathbf{q}+\mathbf{G})
{B_{v\mathbf{k}}^{v^{\prime}\mathbf{k}^{\prime}}}^{*}(\mathbf{q}+\mathbf{G})\\
&\times \Bigg\{
\rz{1-\frac{\hbar\omega_p\left[1-{\epsilon}^{-1}(\mathbf{q}+\mathbf{G},\omega=0)\right]}
{\hbar\omega_p+E_b\left[1-{\epsilon}^{-1}(\mathbf{q}+\mathbf{G},\omega=0)\right]^{1/2}}}\Bigg\}\delta_{\mathbf{q},\mathbf{k}-\mathbf{k}^{\prime}}.
\end{aligned}
\end{equation}
 In their work, \citeauthor{zhang2023effect} examined this dynamical BSE approach against (1) the exact diagonalization approach by solving Eq. \ref{eq:dynamic_BSE} and (2) the perturbation method approach by \citeauthor{rohlfing2000electron} on an organic crystal of naphthalenes on top of the DFT with scissor-shift (DFT+$\Delta$) calculation.
The organic solid shows a particularly strong excitonic effect with the exciton binding energy of $\sim$1.0 eV, and taking into account the dynamical effect for the screened Coulomb interaction results in a noticeable reduction of the excitation energy by $\sim$0.1 eV.
They found that the effective dielectric function method performs very accurately when compared to directly solving the dynamical BSE (Eq. \ref{eq:dynamic_BSE}) through the expensive exact diagonalization method.
\rz{
The frequency dependence of $W$ depends on the plasma frequency, and the validity of this weak-screening limit can be largely gauged by the ratio of the exciton binding energy to the plasma frequency. This approach is therefore expected to become less accurate as this ratio increases. For instance, core-electron excitations, whose binding energies are typically much larger than those of the valence excitons considered here (see, e.g., Ref. \cite{yao2022all}), may provide an insightful case for examining the limitations of this particular effective dielectric function approach in future work.
}

In this work, we develop the 
effective dielectric function approach originally formulated using plane-wave (PW) basis by \citeauthor{zhang2023effect} for our recently-developed all-electron NAO formulation of BSE@$GW$ method for extended systems\cite{zhou2024all}. 
\rz{ In the following section, we first summarize the static BSE formulation in the NAO basis and then introduce how the dynamical screening is incorporated into the BSE kernel by the effective dielectric function.}

\vskip 20 pt

\subsection{Effective Dielectric Function Method with Atom-Centered Orbitals }

\noindent
Throughout this section, we follow the notation with the atom-centered basis such that $i,j$ are used for the Kohn-Sham (KS) orbitals  in the valence band and $a,b$ denote those in the conduction band. For atom-centered orbital (AO) basis, we employ the indices $m$ and $n$, while the Greek letters $\mu, \nu, \alpha,$ and $\beta$ are used for auxiliary basis functions (ABFs).


\vskip 20 pt
\noindent
\textbf{Static BSE in NAO basis:}
For completeness, let us start with our static BSE method formulated using the resolution-of-identity technique for extended systems as detailed in our recent work (Ref. \cite{zhou2024all}).
The products of Bloch-adapted AOs can be expanded using Bloch-adapted atom-centered Auxiliary Basis Functions (ABFs),

\begin{equation}\label{eq:def_aux}
\varphi_{m}^{\mathbf{k}*}(\mathbf{r})\varphi_{n}^{\mathbf{k}^{\prime}}(\mathbf{r})=\sum_{\mu}^{N_{\rm aux}}C_{m,n}^{\mu}(\mathbf{k},\mathbf{k}^{\prime})P_{\mu}^{\mathbf{q}*}(\mathbf{r}).
\end{equation}
where $N_{\rm aux}$ is the number of ABFs within each unit cell and $\mathbf{q}=\mathbf{k}-\mathbf{k}^{\prime}$ is BZ sampling grid for ABFs.
The Bloch-adapted atom-centered ABFs,  $P_{\mu}^{\mathbf{q}}(\mathbf{r})$, are defined through Bloch theorem as \cite{bloch1929quantenmechanik}

\begin{equation}
P_{\mu}^{\mathbf{q}}(\mathbf{r})=\sum_{\mathbf{R}}P_{\mu}(\mathbf{r-R-\tau_{\mu}})e^{i\mathbf{q}\cdot \mathbf{R}}.
\end{equation} 
\noindent
In Eq. \ref{eq:def_aux}, $C_{m,n}^{\mu}(\mathbf{k},\mathbf{k}^{\prime})$ is the AO-based resolution-of-identity (RI) expansion coefficient, which depends on two independent wave-vectors $\mathbf{k}$ and $\mathbf{k}^{\prime}$.
Following Ref. \citenum{ren2012resolution} and \citenum{ren2021all},
the matrix representation of the Coulomb operator $\hat{V}$ and static screened Coulomb operator in terms of the ABFs are 
\begin{equation}\label{eq:coulmb_aux}
\begin{split}
        V_{\mu\nu}(\mathbf{q})&=\iint\frac{P_{\mu}^{\mathbf{q*}}(\mathbf{r}){P}_{\nu}^{\mathbf{q}}(\mathbf{r}^{\prime})}{|\mathbf{r}-\mathbf{r}^{\prime}|}d \mathbf{r} d \mathbf{r}^{\prime}\\ W_{\mu\nu}\,(\mathbf{q})&=\iint P_{\mu}^{\mathbf{q}\ast}(\mathbf{r})\hat{W}(\mathbf{r},\mathbf{r}^{\prime})P_{\nu}^{\mathbf{q}}(\mathbf{r}^{\prime})d \mathbf{r} d \mathbf{r}^{\prime}.
\end{split}
\end{equation}
Based on the definition of the screened Coulomb operator $\hat{W}$, the static screened Coulomb matrix is given by

\begin{equation}\label{eq:screen_aux}
        W_{\mu\nu}(\mathbf{q})=\sum_{\alpha \beta}V_{\mu \alpha}^{\frac{1}{2}}(\mathbf{q})\epsilon_{\alpha\beta}^{-1}(\mathbf{q};\omega=0)V_{ \beta \nu}^{\frac{1}{2}}(\mathbf{q})
\end{equation}
where $V^{\frac{1}{2}}$ represents the square root of the Coulomb matrix $V$ and $\epsilon$ represents the symmetrized static dielectric function, whose matrix elements are computed as 
\begin{equation}\label{eq:dielectric_aux}
    \epsilon_{\mu\nu}(\mathbf{q};\omega=0)=\delta_{\mu\nu}-\sum_{\alpha\beta}V_{\mu \alpha}^{\frac{1}{2}}(\mathbf{q})\chi_{0,\alpha\beta}(\mathbf{q};\omega=0)V_{ \beta \nu}^{\frac{1}{2}}(\mathbf{q}).
\end{equation}
Within the random phase approximation (RPA) expression, $\chi_0$ is the non-interacting static response function, according to the Adler-Wiser formula\cite{adler1962quantum,wiser1963dielectric}. The detailed procedure for calculating the function $\chi_{0,\alpha\beta}$ within the ABF basis can be found in Refs. \citenum{ren2021all} and \citenum{zhou2024all}. 
Similarly to Eq.\ref{eq:def_aux}, the product of two Bloch-adapted KS orbitals
 can be expressed using the RI expansion coefficients $\tilde{C}(\mathbf{k},\mathbf{k}^{\prime})$ as
\begin{equation}\label{eq:def_aux_mo}
\psi_{i/a, \mathbf{k}}^{*}(\mathbf{r})\psi_{j/b, \mathbf{k}^{\prime}}(\mathbf{r})=\sum_{\mu}^{N_{\rm aux}}\tilde{C}_{i/a,j/b}^{\mu}(\mathbf{k},\mathbf{k}^{\prime})P_{\mu}^{\mathbf{q}*}(\mathbf{r}).
\end{equation}
The KS-based expansion coefficients are related to the AO-based expansion coefficients (Eq. \ref{eq:def_aux}) through

\begin{equation}\label{eq:def_mo_coeff}
\begin{aligned}
      \tilde{C}_{i,j}^{\mu}(\mathbf{k},\mathbf{k}^{\prime})&=\sum_{m,n}c^*_{m,i}(\mathbf{k})c_{n,j}(\mathbf{k}^{\prime}) C_{m,n}^{\mu}(\mathbf{k},\mathbf{k}^{\prime})  \\
      \tilde{C}_{a,b}^{\mu}(\mathbf{k},\mathbf{k}^{\prime})&=\sum_{m,n}c^*_{m,a}(\mathbf{k})c_{n,b}(\mathbf{k}^{\prime}) C_{m,n}^{\mu}(\mathbf{k},\mathbf{k}^{\prime})  \\      
\end{aligned}
\end{equation}
where $c_{m,i/a}$ are Kohn-Sham orbital coefficients expressed in terms of the Bloch-adapted AO functions. 
The matrix elements of the Coulomb operator $\hat{V}$ and the static screened Coulomb operator $\hat{W}$ needed for constructing the BSE Hamiltonian are expressed as
\begin{equation} \label{eq:eq_V2}
\begin{split}
\braket{i\mathbf{k} a\mathbf{k}|\hat{V}|j \mathbf{k}^{\prime} b\mathbf{k}^{\prime} }
= & \iint d\mathbf{r} d\mathbf{r'} \psi_{i,\mathbf{k}}(\mathbf{r}) \psi_{a,\mathbf{k}}^*(\mathbf{r})v(\mathbf{r},\mathbf{r'}) \psi_{j,\mathbf{k}^{\prime}}^*(\mathbf{r'}) \psi_{b,\mathbf{k}^{\prime}}(\mathbf{r'}) \\
= & \sum_{\mu\nu} \tilde{C}_{i,a}^{\mu*}(\mathbf{k},\mathbf{k}) V_{\mu\nu}(\mathbf{q}=0) \tilde{C}_{j,b}^{\nu}(\mathbf{k}^{\prime},\mathbf{k}^{\prime})
\end{split} 
\end{equation}
\begin{equation} \label{eq:eq_W2}
\begin{split}
\braket{ i\mathbf{k} j\mathbf{k}^{\prime}|\hat{W}|a\mathbf{k}b\mathbf{k}^{\prime}}
= & \iint d\mathbf{r} d\mathbf{r'} \psi_{i,\mathbf{k}}(\mathbf{r}) \psi_{j,\mathbf{k}^{\prime}}^*(\mathbf{r})W(\mathbf{r},\mathbf{r'}) \psi_{a,\mathbf{k}}^*(\mathbf{r'}) \psi_{b,\mathbf{k}^{\prime}}(\mathbf{r'}) \\
= & \sum_{\mu\nu} \tilde{C}_{i,j}^{\mu*}(\mathbf{k},\mathbf{k}^{\prime}) W_{\mu\nu}(\mathbf{q}=\mathbf{k}^{\prime}-\mathbf{k}) \tilde{C}_{a,b}^{\nu}(\mathbf{k},\mathbf{k}^{\prime}).
\end{split} 
\end{equation}
Thus,  the number of $\mathbf{q}$-points available in the calculation is given by the number of $\mathbf{k}$-points used for the BZ sampling.
In practice, the explicit $\mathbf{q}$ dependence of the screened Coulomb interaction $W_{\mu\nu}(\mathbf{q})$ (Eq. \ref{eq:coulmb_aux}-Eq. \ref{eq:dielectric_aux}) represents a significant computational bottleneck, as converging the BSE absorption spectrum typically requires a very dense sampling of BZ. 

 To achieve computational efficiency necessary for applications, this implementation also utilizes the localized resolution-of-identity (LRI) approximation~\cite{ihrig2015accurate} as discussed in our prior works~\cite{ren2021all, zhou2024all}.
Without discussing the details here, an important point of this LRI approach is that the AO-based expansion coefficients (Eq. \ref{eq:def_aux}) 
become dependent of only a single vector, either $\mathbf{k}^{\prime}$ or $\mathbf{k}$ within the BZ, rather than of both simultaneously. As a result, the computational cost and memory storage requirements associated with the RI coefficients can be significantly reduced.
\vskip 20 pt
\noindent
\textbf{\rz{Dynamical BSE with Effective Dielectric Function in NAO basis:} }
\rz{
In the following, we adapt the NAO basis approach developed for the standard static BSE \cite{zhou2024all} for the dynamical BSE with the effective dielectric function.
}
In the standard reciprocal lattice vector representation, the matrix elements of the static screened Coulomb interaction (Eq.~\ref{eq:eq_W2}) are given by
\begin{align}
      \braket{ i\mathbf{k} j\mathbf{k}^{\prime}|\hat{W}(\omega=0)|a\mathbf{k}b\mathbf{k}^{\prime}}=&\frac{1}{\Omega}\sum_{\mathbf{q},\mathbf{G}} v(|\mathbf{q}+\mathbf{G}|)\epsilon^{-1}(\mathbf{q}+\mathbf{G},\omega=0)   \nonumber \\\times& B_{i\mathbf{k}}^{j\mathbf{k}^{\prime}}(\mathbf{q}+\mathbf{G})B_{a\mathbf{k}}^{b\mathbf{k}^{\prime}*}(\mathbf{q}+\mathbf{G}).   
      \label{eq:static_interaction_plane_wave}
\end{align}
Here, $B_{i\mathbf{k}}^{j\mathbf{k}^{\prime}}(\mathbf{q}+\mathbf{G})$ and  $B_{a\mathbf{k}}^{b\mathbf{k}^{\prime}*}(\mathbf{q}+\mathbf{G})$ are Bloch integrals as defined in Eq. \ref{eq:w_eff}, and $W(\mathbf{q}+\mathbf{G}, \omega=0) = v(|\mathbf{q}+\mathbf{G}|) \epsilon^{-1}(\mathbf{q}+\mathbf{G}, \omega=0)$ gives the static screened Coulomb interaction.
 %
As discussed above, the dynamical BSE approach\cite{zhang2023effect} by \citeauthor{zhang2023effect} captures the frequency dependence of the screened interaction through an effective dielectric function as seen in Eq. \ref{eq:w_zhang}, and the effective dielectric matrix is
\begin{equation}\label{effective_inverse_dielectric_def}
\epsilon^{-1}_{\mathrm{eff}}(\mathbf{q}+\mathbf{G}) = 1 - \rz{\frac{\hbar \omega_p \left[1-\epsilon^{-1}(\mathbf{q}+\mathbf{G},\omega=0)\right] }{\hbar \omega_p + E_b\left[1-\epsilon^{-1}(\mathbf{q}+\mathbf{G},\omega=0)\right]^{1/2}}},
\end{equation}
where $\omega_p$ is the plasma frequency and $E_b$ is the binding energy of the specific exciton under consideration. 
Thus, the effective screened Coulomb interaction, $W^{\rm eff}(\mathbf{q}+\mathbf{G})=v(|\mathbf{q}+\mathbf{G}|) \epsilon^{-1}_{\rm eff}(\mathbf{q}+\mathbf{G}) $, can be obtained once the static dielectric matrix $\epsilon(\omega=0)$, the bare Coulomb matrix $V$, the plasma frequency, and the exciton binding energy are known. 
Notably, in the limit of the exciton binding energy $E_b$ approaching zero, the effective inverse dielectric function $\epsilon^{-1}_{\mathrm{eff}}$ reduces to the static inverse dielectric function $\epsilon^{-1}(\omega=0)$ used in conventional static BSE calculations (Eq.~\ref{eq:static_interaction_plane_wave}).


While the PW basis representation naturally caters to the standard reciprocal lattice vector expressions, our all-electron NAO formulation requires a different treatment in the numerical implementation.  
We utilize the localized resolution-of-identity (LRI) approximation \cite{ren2012resolution,ihrig2015accurate,ren2021all}, and the required operators are represented in auxiliary basis functions (ABFs) (see Eq. \ref{eq:def_aux}). 
Analogously to the static BSE implementation, the effective screened Coulomb interaction in the ABF representation is

\begin{equation}\label{eq:screen_aux_eff}
W^{\mathrm{eff}}_{\mu\nu}(\mathbf{q}) = \sum_{\alpha \beta}V_{\mu \alpha}^{\frac{1}{2}}(\mathbf{q})[\epsilon_{\mathrm{eff}}^{-1}(\mathbf{q})]_{\alpha\beta}V_{ \beta \nu}^{\frac{1}{2}}(\mathbf{q})
\end{equation}
where $\left[\epsilon_{\mathrm{eff}}^{-1}(\mathbf{q})\right]_{\alpha\beta}$ denotes the matrix representation of the effective inverse dielectric function in the auxiliary atom-centered basis. 

Unlike for the conventional diagonal approximation employed  with the plane-wave formulation (Eq. \ref{eq:w_zhang}), the dielectric matrix evaluated in the ABFs is inherently dense and non-diagonal. 
Since the effective screening model (Eq.~\ref{effective_inverse_dielectric_def}) involves a nonlinear functional transformation of the dielectric response, we evaluate it in the eigenbasis of the static inverse dielectric matrix where the screening channels (different eigenspaces) become decoupled. 
We begin by performing a spectral decomposition of the static inverse dielectric matrix in the auxiliary basis
\begin{equation}\label{eq:spectral_decomp}
     \epsilon_{\mu\nu}^{-1} (\mathbf{q},\omega=0) = \sum_{\lambda}X_{\mu\lambda}(\mathbf{q})\widetilde{\epsilon}^{-1}_{\lambda}(\mathbf{q},\omega=0)X^*_{\nu\lambda}(\mathbf{q})
\end{equation}
where $X$ is the unitary transformation matrix whose columns are the eigenvectors of $\epsilon_{\mu\nu}^{-1} (\omega=0)$, and $\widetilde{\epsilon}^{-1}_{\lambda}(\omega=0)$ are its corresponding eigenvalues.  
\rz{All $N_{\rm aux}$ eigenvalues are retained in Eq.~\ref{eq:spectral_decomp}, and we apply no truncating approximation. 
The eigenmodes whose $\widetilde{\epsilon}^{-1}_{\lambda}$ lies close to unity are barely screened and contribute little to $W^{\rm eff}$, so discarding them would reduce the cost of the back-transformation in Eq.~\ref{eq:inverse_dielectric_eff}. 
Such a truncation would also change how the spectral decomposition itself is carried out. 
We currently obtain Eq.~\ref{eq:spectral_decomp} by diagonalizing the dense $N_{\rm aux} \times N_{\rm aux}$ matrix in full, but if only the strongly screened eigenmodes are needed, an iterative eigensolver of the Davidson or Lanczos type can return the truncated eigen-space alone without constructing the rest of the spectrum. 
We leave the development of such truncation approaches and the error analysis  for the future work.
}

Following the spectral decomposition, the nonlinear transformation (Eq.~\ref{effective_inverse_dielectric_def}) is applied directly to the eigenvalues. 
The effective inverse dielectric matrix is subsequently constructed as

\begin{equation}\label{eq:dynanmical_eigenvalue}
(\widetilde{\epsilon}^{-1}_{\mathrm{eff}})_{\lambda}(\mathbf{q}) = 1 - \frac{\hbar \omega_p [1-\widetilde{\epsilon}^{-1}_{\lambda}(\mathbf{q},\omega=0)] }{\hbar \omega_p + E_b[1-\widetilde{\epsilon}^{-1}_{\lambda}(\mathbf{q},\omega=0)]^{\frac{1}{2}}}. 
\end{equation}
\rz{Eq. \ref{eq:dynanmical_eigenvalue} is  algebraically equivalent to Eq. \ref{eq:w_zhang} and Eq. \ref{effective_inverse_dielectric_def}, differing only in whether the screening channel is labeled by a plane-wave component $\mathbf{q}+\mathbf{G}$ or by an eigenvalue index $\lambda$ of the dielectric matrix in the auxiliary basis.} 
Finally, this resulting matrix is transformed back from the eigenspace into the auxiliary basis as
\begin{equation}\label{eq:inverse_dielectric_eff}
    [\epsilon_{\mathrm{eff}}^{-1}(\mathbf{q})]_{\alpha\beta} = \sum_{\lambda}X_{\alpha\lambda}(\mathbf{q})(\widetilde{\epsilon}^{-1}_{\mathrm{eff}})_{\lambda}(\mathbf{q}) X^*_{\beta\lambda}(\mathbf{q}).
\end{equation}
This effective inverse dielectric matrix in the auxiliary basis then enters Eq.~\ref{eq:screen_aux_eff} to compute the screened interaction.
\rz{The dielectric matrix $\epsilon_{\mu\nu}(\mathbf{q})$ (see  Eq.~\ref{eq:dielectric_aux}) is inverted in full at every $\mathbf{q}$ point and is never truncated to its diagonal, so the screened interaction of Eq.~\ref{eq:screen_aux_eff} carries the local-field effects by construction. 
Unlike for the traditional plane-wave basis, the compact nature of the atom-centered basis representation allows us to perform the full inversion without resorting to 
the diagonal approximation ($\mathbf{G} = \mathbf{G}^{\prime}$), which is widely used in the plane-wave formulation of Eq.~\ref{eq:w_zhang}.}

\subsection{Symmetry Adaptation Irreducible  Brillouin Zone (IBZ) Mapping }
Evaluating the screened Coulomb interaction $W_{\mu\nu}(\mathbf{q})$ over a dense $\mathbf{q}$-point grid is necessary for converging BSE for optical absorption spectrum in practical calculations\cite{zhou2024all} but computationally expensive.
In the case of static BSE calculations, this requires the evaluation of the RPA polarizability $\chi_0(\mathbf{q})$ and the static dielectric function $\epsilon(\mathbf{q})$ (Eq.~\ref{eq:dielectric_aux}) on a dense $\mathbf{q}$-point grid. 
Also, as we move beyond the static dielectric function for dynamical BSE, the construction of the effective dielectric matrix $W^{\mathrm{eff}}_{\mu\nu} (\mathbf{q})$ (Eqs.~\ref{eq:screen_aux_eff}--\ref{eq:inverse_dielectric_eff}) requires an additional matrix diagonalization for each $\mathbf{q}$-point to obtain the effective inverse dielectric matrix $\epsilon_{\mathrm{eff}}^{-1}(\mathbf{q})$. 
Consequently, the explicit $\mathbf{q}$ dependence of the dielectric response remains the central bottleneck for both static and dynamical BSE calculations.
In contrast, the matrix operations required to rotate a tensor from one $\mathbf{q}$-point to another are computationally negligible. 
Therefore, we propose to calculate the static dielectric function and the corresponding effective screened interaction only at $\mathbf{q}$ points within the irreducible Brillouin zone (IBZ) and map onto the full BZ as discussed in the following. 
These matrices at arbitrary $\mathbf{q}$-points in the full BZ are then recovered through crystallographic symmetry operations. It is worth noting that while this approach offers an added advantage for the dynamical BSE where the screening is more complex, it is equally efficient and applicable to the standard static BSE.

Consider a space group symmetry operation $\hat{S} = \{R|\mathbf{f}\}$, consisting of a rotation $R$ and a fractional translation $\mathbf{f}$. An arbitrary vector $\mathbf{q}$ in the full BZ is related to an irreducible point $\mathbf{q}_{\mathrm{IBZ}}$ via $\mathbf{q} = \hat{S} \mathbf{q}_{\mathrm{IBZ}} + \mathbf{G}$, where $\mathbf{G}$ is a reciprocal lattice vector. Under this $\hat{S}$ operation, the Bloch-adapted ABFs transform as
\begin{equation}
    \hat{S} P_{\nu}^{\mathbf{q}_{\mathrm{IBZ}}}(\mathbf{r}) = \sum_{\mu} \mathcal{D}_{\mu\nu}(\hat{S}, \mathbf{q}) P_{\mu}^{\mathbf{q}}(\mathbf{r}).
\end{equation}
where the transformation matrix $\mathcal{D}_{\mu\nu}$ accounts for the atom-centered nature of the basis functions. The composite index $\mu$ is associated with an atom $I$ located at $\boldsymbol{\tau}_I$, an angular momentum $l$, and a magnetic quantum number $m$ (denoted collectively as $\mu \equiv \{I, l, m\}$). The matrix element $\mathcal{D}_{\mu\nu}$ is constructed as a product of three components
\begin{equation}
    \mathcal{D}_{\mu\nu}(\hat{S}, \mathbf{q}) = \delta_{I, \hat{S}(J)} \delta_{l l'} 
    \mathcal{R}_{m m'}^{l}(R) e^{-i \mathbf{q} \cdot \mathbf{R}_{\mathrm{shift}}}.
\end{equation}
Here, $\nu \equiv \{J, l', m'\}$ is the index of the untransformed basis function. The components are defined as follows:
\begin{enumerate}
    \item \textbf{Atom Permutation} ($\delta_{I, \hat{S}(J)}$): The operation maps the atom $J$ to atom $I$ (or its periodic image). The matrix element is non-zero only if the basis functions are centered on symmetry-equivalent atoms.
    \item \textbf{Geometric Rotation} ($\mathcal{R}_{m m'}^{l}$): This is the real Wigner rotation matrix that mixes the magnetic quantum numbers $m'$ into $m$ for the specific angular momentum $l$ under the rotation $R$.
    \item \textbf{Bloch Phase Factor} ($e^{-i \mathbf{q} \cdot \mathbf{R}_{\mathrm{shift}}}$): Since the symmetry operation may map an atom onto a neighboring unit cell, a phase shift is required. The lattice shift vector is defined as $\mathbf{R}_{\mathrm{shift}} = R\boldsymbol{\tau}_J + \mathbf{f} - \boldsymbol{\tau}_I$. 
    
\end{enumerate}
This approach was also used in Ref. \citenum{cao2025applying} in the context of hybrid exchange-correlation functional in DFT calculations. 
In the context of BSE calculation here, the effective screened Coulomb interaction matrix at the arbitrary $\mathbf{q}$ in the full BZ is obtained via 
\begin{equation}\label{eq:W_symmetry_map}
    W^{\mathrm{eff}}_{\mu\nu}(\mathbf{q}) = \sum_{\alpha\beta} \mathcal{D}_{\mu\alpha}(\hat{S}, \mathbf{q}) W^{\mathrm{eff}}_{\alpha\beta}(\mathbf{q}_{\mathrm{IBZ}}) \mathcal{D}^*_{\nu\beta}(\hat{S}, \mathbf{q}).
\end{equation}
By replacing the explicit calculation of $\epsilon^{-1}$ with this algebraic mapping, we avoid redundant expensive computations for symmetry-equivalent $\mathbf{q}$-points, significantly accelerating the construction of both the static and dynamical BSE Hamiltonian matrices. \rz{Eq.~\ref{eq:W_symmetry_map} is an exact algebraic identity rather than an approximation, since $\mathcal{D}(\hat{S},\mathbf{q})$ is unitary.
The same identity covers the dynamical case without modification. $W^{\mathrm{eff}}_{\mu\nu}(\mathbf{q})$ is obtained by applying the effective screening model to the eigenvalues of $\epsilon^{-1}_{\mu\nu}(\mathbf{q},\omega=0)$ (Eq.~\ref{eq:spectral_decomp}), and a unitary transformation leaves those eigenvalues unchanged, so the mapping and the construction of $W^{\mathrm{eff}}$ commute.}
Application of this technique is employed for the computationally intensive BSE@$G_0W_0$ calculations and discussed in the following section.

\section{Results and Discussion}

We implemented the above effective dielectric function method for dynamical BSE in FHI-aims code\cite{blum2009ab,abbott2025roadmap}.
As done in the work by \citeauthor{zhang2023effect}\cite{zhang2023effect}, our NAO-based implementation is demonstrated here for crystalline naphthalene, which is an organic crystal with a large exciton binding energy of approximately 1 eV. \rz{This large binding energy makes this organic crystal a particularly informative test case since the dynamical correction is then expected to be significant enough for comparing the NAO and plane-wave formulations.}
Our NAO-based implementation is first compared to the PW-PAW implementation by \citeauthor{zhang2023effect}\cite{zhang2023effect} in BSE@DFT+$\Delta$ calculation.  
We then demonstrate  BSE@$G_0W_0$ calculation by exploiting the IBZ-to-full-BZ mapping method discussed above to overcome the large computational cost associated with the BZ integration. 
In BSE@$G_0W_0$ calculations, the $\mathbf{q}$-point grid of the screened Coulomb matrix (see e.g. Eq. \ref{eq:eq_W2} and Eq.\ref{eq:screen_aux_eff}) must be commensurate with the $\mathbf{k}$-point grid used for the BZ sampling in $GW$ calculation while a dense $\mathbf{q}$-point grid is necessary for converging the optical absorption spectrum in BSE calculations\cite{zhou2024all}.

\begin{figure}[htbp]
\centering
\includegraphics[width=0.75\linewidth]{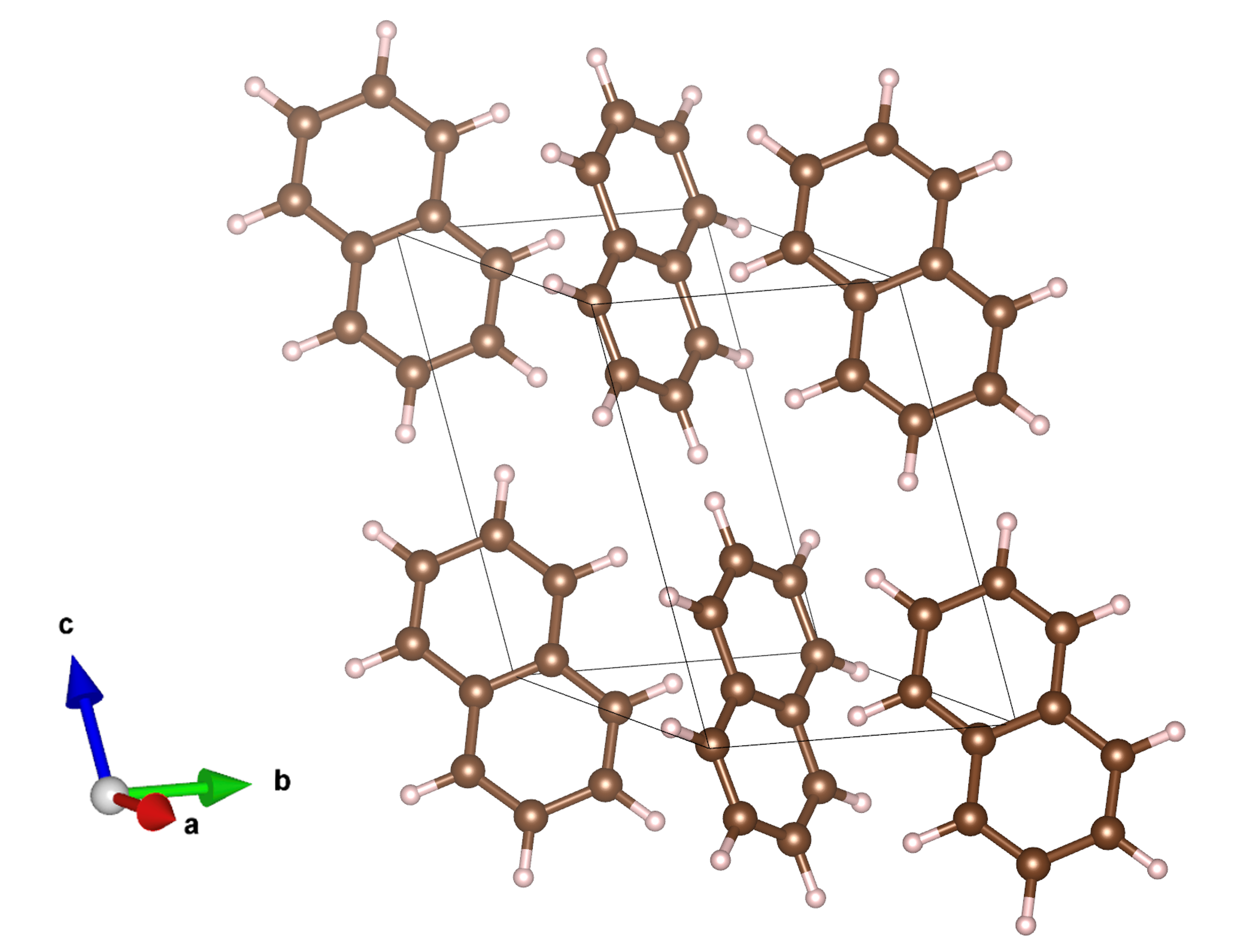}
\caption{
Crystal structure of monoclinic naphthalene ($\rm C_{10}H_{8}$). Carbon and hydrogen atoms are shown as brown and white spheres, respectively. The unit cell contains two conjugated naphthalene molecules arranged in a herringbone-like packing motif. Lattice vectors a, b, and c are indicated in the figure.
}
\label{fig:struc}
\end{figure}

\subsection{Comparison of all-electron NAO implementation to PW-PAW implementation for BSE@DFT+$\Delta$}
We used the same crystal structure coordinates as \citeauthor{zhang2023effect}, which were optimized using the DFT-D2 method of Grimme\cite{grimme2006semiempirical} to account for van der Waals interaction\cite{zhang2023effect}. 
We first performed DFT calculation using Perdew, Burke, and Ernzerhof (PBE) generalized gradient approximation (GGA) functional \cite{perdew1996generalized}, using the Tier 2 NAO basis set\cite{blum2009ab}. 
The simulation cell includes 36 atoms and 136 electrons, and the Brillouin zone integration was performed with the $\Gamma$-centered sampling of 5$\times$7$\times$5 $\mathbf{k}$-points.
We then apply the ``scissor'' operator such that all the conduction band KS eigenvalues are constantly shifted by 1.55 eV as done in Ref. \citenum{zhang2023effect}, as often referred to as DFT+$\Delta$ method. 
The macroscopic dielectric function $\varepsilon_M(\omega)$ is given by $\varepsilon_M(\omega) = {1}/{\lim_{\mathbf{q}\to 0} \rz{\epsilon^{-1}(\mathbf{q}, \omega)}}$. 
Although the f-sum rule, $\int d\omega\ \omega {\rm Im} \epsilon_M^{-1}=-\frac{\pi}{2}\omega_p^2,$ can be used to obtain the plasma frequency, 
we use the reported value of 17.9 eV from Ref. \citenum{zhang2023effect} here for this benchmarking comparison of our NAO-based implementation. 
\citeauthor{zhang2023effect} discuss the negligible dependence of the BSE optical absorption spectrum on the numerical precision associated with computing the plasma frequency\cite{zhang2023effect}.
Because the lowest eigenstate (the first excited state) for this system is essentially \textit{dark} (i.e. negligible oscillator strength), we use the peak of the absorption spectrum for defining the exciton binding energy ($E_{b}=E_{gap}^{QP}-E_{gap}^{opt}$) as done also by \citeauthor{zhang2023effect}.
Table \ref{table:bse_dft} summarizes the comparison.

\begin{table}[H]
\centering
\begin{tabular}{c|c|c|c|c}
\hline 
\textbf{Ref. \citenum{zhang2023effect}} & KS Gap (eV) & Optical Gap (eV) & Computed BE (eV) & Input BE (eV)   \\ \hline
\textit{DFT+$\Delta$ }   &  4.78  &   NA         &     NA   & NA   \\ 
\textit{sta. BSE  }   & NA   &    3.78        &   1.00      & NA \\ 
\textit{dyn. BSE } & NA   &   3.68          &  1.10       & 1.00   \\ 
\textit{dyn. BSE } & NA   &   3.67          &  1.11      &  1.14  \\ \hline

\hline \hline
\textbf{This work} & KS Gap (eV) & Optical Gap (eV) & Computed BE (eV) & Input BE (eV)   \\ \hline
\textit{DFT+$\Delta$ }     &  4.78  &     NA       &   NA    &   NA    \\ 
\textit{sta. BSE  }     &  NA   &   3.81          &     0.97    &  NA  \\ 
\textit{dyn. BSE } &  NA   &     3.68          &   1.10   &   1.06 (Ref. \citenum{zhang2023effect})    \\ \hline
\end{tabular}
\caption{Comparison of the all-electron NAO-based implementation of the effective dielectric function method for dynamical BSE@DFT+$\Delta$ to the reference work based on plane-wave-based implementation by \citeauthor{zhang2023effect}\cite{zhang2023effect}. BE stands for the exciton binding energy. }
\label{table:bse_dft}
\end{table}

\begin{figure}[htbp]
\centering
\includegraphics[width=0.85\linewidth]{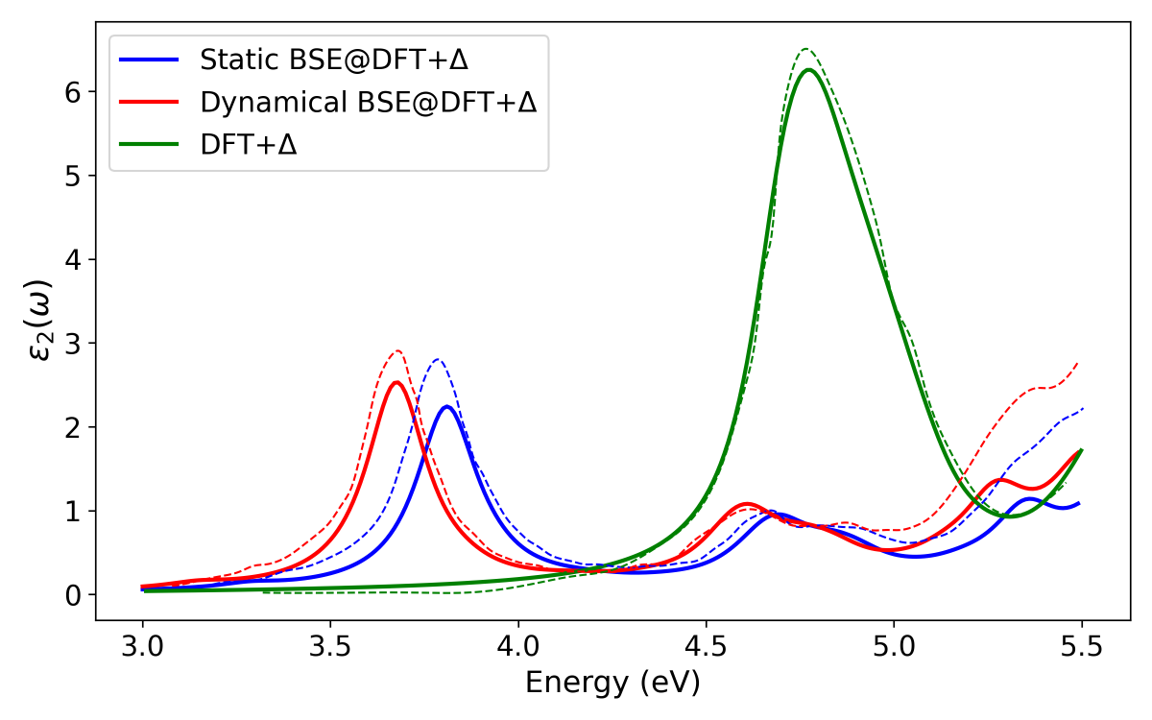}

\caption{
Comparison of the imaginary part of the dielectric function 
(i.e. optical absorption spectrum) in the $b$ direction (see Fig. 1), 
using DFT+$\Delta$, static BSE@DFT+$\Delta$, and 
dynamical BSE@DFT+$\Delta$. 
The input exciton binding energy of $1.06\,\mathrm{eV}$ and the plasma frequency 
of $17.9\,\mathrm{eV}$ are used in all the calculations as done in the reference work. 
\rz{The dashed lines are the reference result of \citeauthor{zhang2023effect}\cite{zhang2023effect}, obtained by digitizing the spectra reported in that work.}
}
\label{fig:bse_dft}
\end{figure}

In the BSE@DFT+$\Delta$ approach without $GW$ calculation, 
an accurate estimate of the exciton binding energy is not available.
In the original work by \citeauthor{zhang2023effect}\cite{zhang2023effect}, the dependence on the input exciton binding energy (BE) was discussed extensively. 
As shown in Table \ref{table:bse_dft}, for example, they found that the computed optical gap from the dynamical BSE calculation changes only by 0.01 eV, regardless of using 1.00 or 1.14 eV for the input exciton BE.  
For our calculation, we use the input exciton BE of 1.06 eV as done in Ref. \citenum{zhang2023effect}.
Five valence and eight conduction band states are included in solving BSE, covering the same energy window of 
9 eV as the referenced work\cite{zhang2023effect}.
The optical absorption spectrum computed from the static and dynamical BSE calculations is shown in Figure \ref{fig:bse_dft}.
Our all-electron NAO-based implementation shows an excellent agreement with the original PW-based implementation within 30 meV for both
static and dynamical BSE calculations. 
Specifically, our results (solid lines) show excellent agreement with the reference PW results (dashed lines) in terms of both peak positions and intensities. 
We attribute the minor discrepancy in peak intensity to the calculation of the transition dipole moments, as such a difference is also present even for the DFT calculation (green line).

\subsection{Performance of IBZ  mapping for $W(\mathbf{q})$ }
\begin{figure}[htbp]
\centering
\includegraphics[width=0.85\linewidth]{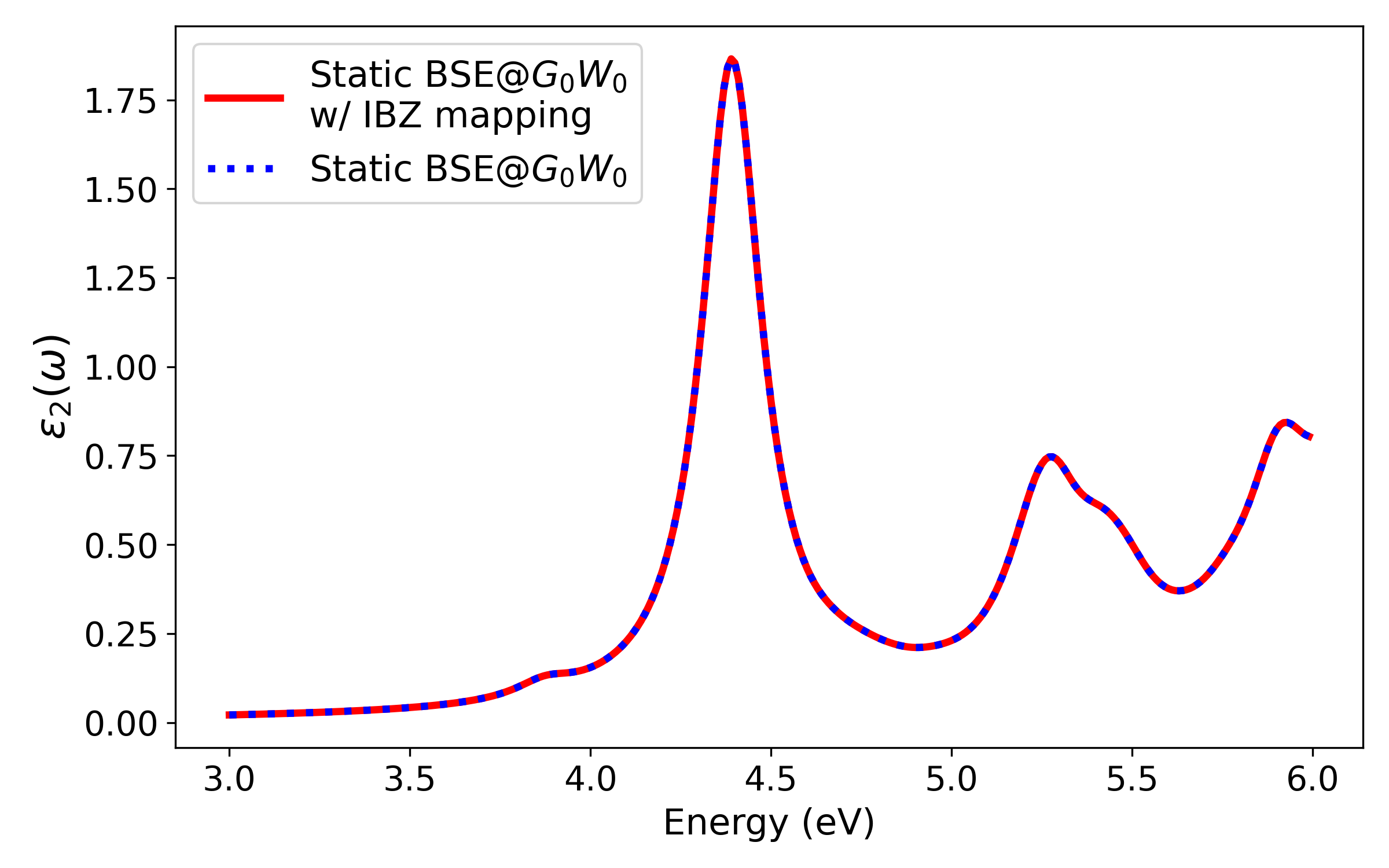}
\caption{
Comparison of the imaginary part of the dielectric function 
(i.e. optical absorption spectrum) in the $b$ direction (see Fig. 1), computed using static BSE@$G_0W_0$ method, with and without the IBZ mapping scheme. 
In both cases, the BSE calculations utilize a $5\times7\times5$ $\mathbf{k}$-point grid. 
Only 52 $\mathbf{q}$-points are needed with the IBZ mapping method while 175 $\mathbf{q}$-points are sampled in the full BZ (see Table \ref{table:bz}). }
\label{fig:IBZ_compare}
\end{figure}

As we focus next on the computationally expensive BSE@$GW$ demonstration of the dynamical BSE method, the IBZ mapping approach, as described in the Theoretical Methods section, is employed.
To validate the accuracy of this symmetry-adapted IBZ mapping, we compared the optical absorption spectra obtained from the standard BSE@$GW$ using the full BZ sampling with that using the IBZ mapping scheme. 
As shown in Figure~\ref{fig:IBZ_compare}, both approaches yield identical results.
The computational advantages of this strategy are highlighted by the $\mathbf{q}$-point counts presented in Table~\ref{table:bz}. 
The IBZ mapping significantly reduces the computational cost of calculating the screened Coulomb interaction (i.e.  $W_{\mu\nu}(\mathbf{q},\omega=0)$ and $W^{\rm eff}_{\mu \nu}(\mathbf{q})$), and the benefit becomes increasingly more pronounced as the BZ sampling points increase. 
For crystalline naphthalene (Figure~\ref{fig:struc}), which belongs to the $C_{2h}$ point group, a $5 \times 7 \times 5$ $\mathbf{q}$-point grid is reduced such that the IBZ contains only $\sim$ 30\% of the points in the full BZ. 
This reduction substantially accelerates both static and dynamical BSE calculations. 
\rz{The savings therefore grow with the symmetry of the crystal. Table S1 of the Supporting Information gives the $\mathbf{q}$-point counts for $\Gamma$-centered $n\times n\times n$ grids across all crystal symmetries.}

\begin{table}[H]
\centering
\caption{Comparison of $\mathbf{q}$-point grids in the irreducible Brillouin zone (IBZ) and the full Brillouin zone (BZ) of naphthalene crystal.  }
\label{table:bz}
\begin{tabular}{|c|c|c|}
\hline 
\textbf{$\mathbf{q}$-point grid} & \textbf{No. in IBZ} & \textbf{No. in Full BZ} \\ \hline
$2\times 3\times 2$ & 8 & 12 \\ 
$3\times 4\times 3$ & 15 & 36 \\ 
$5\times 7\times 5$ & 52 & 175 \\ \hline 
\end{tabular}
\end{table}

\subsection{Demonstration with BSE@$G_0W_0$}
We now demonstrate the applicability of the effective dielectric function approach for dynamical BSE calculation in combination using quasiparticle energies by performing the single-shot $G_0W_0$ calculation. The same optimized crystal structure of naphthalene was employed as in the BSE@DFT+$\Delta$ benchmark discussed above. 
To improve the resolution-of-identity approximation, additional $4f$ spherical 
harmonic functions were included in the construction of the auxiliary basis 
set~\cite{ren2012resolution, ren2021all}. 
For the frequency integration in the self-energy calculation, 
we utilized 60 frequency points within the Padé approximation for 
analytic continuation\cite{zhou2024all}. 

As discussed in our NAO-based BSE development work~\cite{zhou2024all}, the $\mathbf{k}$-point grid required for converging the optical absorption spectrum is significantly denser than that needed for $G_0W_0$ self-energy convergence. To improve computational efficiency, 
we can decouple these two requirements. 
Specifically, while the BSE requires quasiparticle (QP) energies on a dense $\mathbf{k}$-point grid, the self-energy $\Sigma_{n}^{G_0W_0}(\mathbf{k}, i\omega)$ at those $\mathbf{k}$-points can be accurately determined using a much coarser BZ sampling of the screened interaction $W$ on a grid denoted as $\mathbf{k}_0$~\cite{ren2021all},
\begin{equation}
\begin{aligned}\label{eq:GW_Sigma}
\Sigma_{n}^{G_0W_0}(\mathbf{k}, i\omega) &= \iint d\mathbf{r} d\mathbf{r}^{\prime} \psi_{n\mathbf{k}}^*(\mathbf{r}) \Sigma^{G_0W_0}(\mathbf{r}, \mathbf{r}^{\prime}, i\omega) \psi_{n\mathbf{k}}(\mathbf{r}^{\prime}) \\
&= -\frac{1}{2\pi} \sum_{m,\mathbf{k}_{0}} \sum_{\mu,\nu} \int_{-\infty}^\infty d\omega^{\prime} \frac{C_{n,m}^\mu(\mathbf{k}, \mathbf{k}-\mathbf{k}_{0}) W_{\mu\nu}(\mathbf{k}_{0}, i\omega^{\prime}) C_{m,n}^\nu(\mathbf{k}-\mathbf{k}_{0}, \mathbf{k})}{i\omega - i\omega^{\prime} + \mu - \epsilon_{m \mathbf{k}-\mathbf{k}_0}}
\end{aligned}\end{equation}
where $\omega$ is a frequency point on the imaginary axis, $\epsilon_{m \mathbf{k}-\mathbf{k}_0}$ are the underlying KS orbital energies at $\mathbf{k}-\mathbf{k}_0$, and $\mu$ is the electronic chemical potential.  
Here, the screened interaction $W_{\mu\nu}(\mathbf{k}_{0}, i\omega^{\prime})$ is represented in the auxiliary basis and evaluated at the coarser $\mathbf{k}_0$-points.
For the naphthalene crystal, the convergence test in  Table~\ref{table:bse_GW}, comparing $2 \times 3 \times 2$ and $3 \times 4 \times 3$ $\mathbf{k}_0$-point grids for the $G_0W_0$ calculation, shows that the QP band gaps agree within $10$~meV. 
As seen in Table~\ref{table:bse_GW}, the subsequent static (and also dynamical) BSE calculations on the dense $5 \times 7 \times 5$ $\mathbf{k}$-grid yield virtually identical optical gaps regardless of whether the $2 \times 3 \times 2$ or $3 \times 4 \times 3$ $\mathbf{k}_0$-point grids are used in preceding $GW$ calculation, indicating that $W(\mathbf{q}=\mathbf{k}_0)$ is already converged with the coarser $2 \times 3 \times 2$ $\mathbf{k}_0$-point grid. 

After performing $G_0W_0$ calculation, we first perform a static BSE calculation to obtain the optical absorption spectrum and the corresponding optical gap. 
Five valence and eight conduction band states are included in solving BSE.
The exciton binding energy (BE) obtained as the difference between the $G_0W_0$ quasiparticle gap and the static BSE optical gap is 0.98~eV. 
This exciton BE value is then used as the input BE for constructing the effective dielectric function in the dynamical BSE calculation. 
\rz{This input BE value from the static BSE can be replaced by the BE from the dynamical BSE calculation in subsequent calculations. 
The procedure could then be repeated until convergence is achieved in principle. 
For the present system studied here, however, such an iterative procedure is found unnecessary as the dependence on the input BE is quite weak.
The reference calculations in Table \ref{table:bse_dft} show that changing the input BE from 1.00 to 1.14 eV shifts the dynamical BSE optical gap by only 0.01 eV. 
As seen in Table \ref{table:bse_GW} for BSE@$G_0W_0$ calculation, the input BE of 0.98 eV (from the static BSE) is already similarly close to the output BE of 1.10 eV from the dynamical BSE calculation, differing only by 0.12 eV.
Since BSE@$G_0W_0$ shows a sensitivity to the input BE similar to that of BSE@DFT+$\Delta$\cite{zhang2023effect}, the next iteration of the dynamical BSE with the updated BE would change the optical gap by only about 0.01 eV. Performing such an iterative procedure is certainly possible, and it would make a difference for systems in which the dynamical correction to the binding energy is large compared with the scale over which $\epsilon^{-1}_{\rm eff}$ varies with $E_b$ (Eq.~\ref{effective_inverse_dielectric_def}).}

\begin{table}[H]
\centering
\begin{tabular}{c|c|c|c|c}
\hline 
\textbf{$2\times3\times2$} $\mathbf{k}_0$-point& QP Gap (eV) & Optical Gap (eV) & Computed BE (eV) & Input BE (eV)   \\ \hline
\textit{$G_0W_0$ }   &  5.37  &   NA         &     NA   & NA   \\ 
\textit{BSE (static)  }   & NA   &    4.39        &   0.98      & NA \\ 
\textit{BSE (dyn.)} & NA   &   4.27          &  1.10       & 0.98  \\ 
\hline \hline
\textbf{$3\times4\times3$} $\mathbf{k}_0$-point & QP Gap (eV) & Optical Gap (eV) & Computed BE (eV) & Input BE (eV)   \\ \hline
\textit{$G_0W_0$ }   &  5.37  &   NA         &     NA   & NA   \\ 
\textit{BSE (static)  }   & NA   &    4.40        &   0.97      & NA \\ 
\textit{BSE (dyn.)} & NA   &   4.27          &  1.10       & 0.97  \\ 
\hline
\end{tabular}
\caption{Comparison of $G_0W_0$ and BSE results using two different $\mathbf{k}_0$-point grids for the $G_0W_0$ self-energy calculation. The BSE absorption spectra and exciton binding energies (BE) were evaluated on a converged $5 \times 7 \times 5$ $\mathbf{k}$-point grid.}
\label{table:bse_GW}
\end{table}

\begin{figure}[htbp]
\centering
\includegraphics[width=0.85\linewidth]{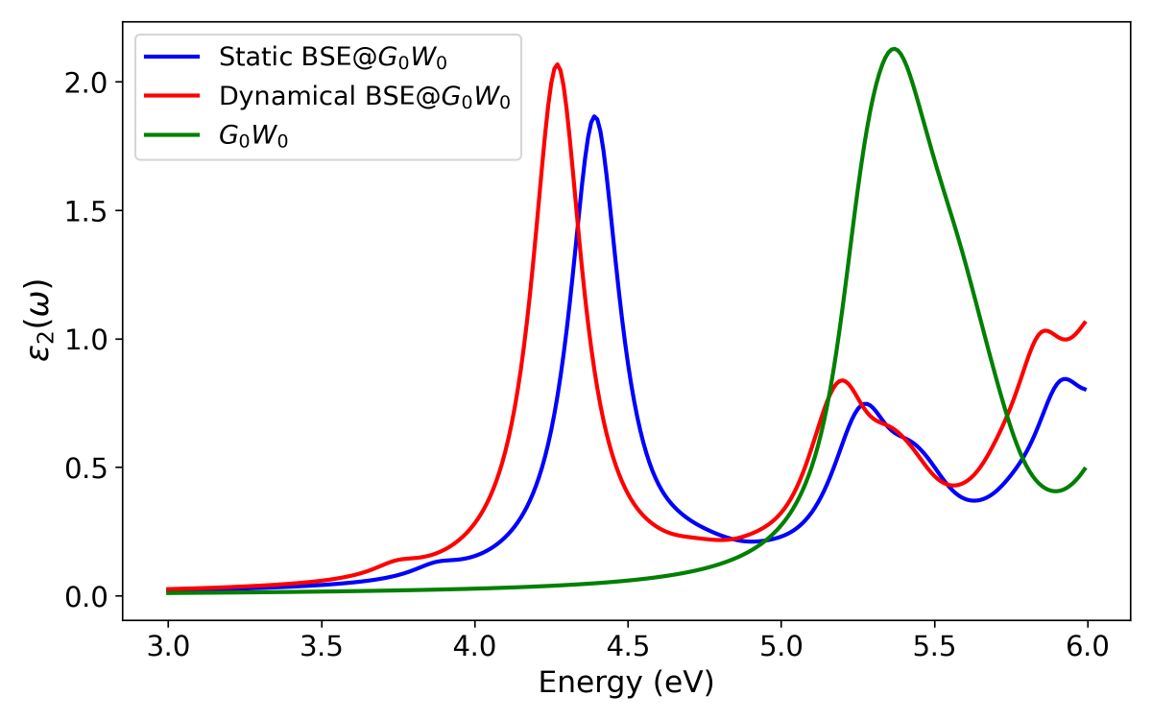}

\caption{
Comparison of the imaginary part of the dielectric function 
(i.e. optical absorption spectrum) in the $b$ direction (see Fig. 1),  
using $G_0W_0$, static BSE@$G_0W_0$, and dynamical BSE@$G_0W_0$ calculations. 
$G_0W_0$ calculation uses the $\Gamma$-centered $2\times3\times2$ $\mathbf{k}_0$-point grid for the calculation of $W(\mathbf{q})$ and obtains the quasiparticle (QP) energies on the denser $5\times7\times5$ $\textbf{k}$-point grid
(Eq. \ref{eq:GW_Sigma}). 
Dynamical BSE@$G_0W_0$ calculation uses the plasma frequency of 17.9 eV and the input exciton binding energy of 0.98 eV from static BSE@$G_0W_0$ calculation.
\rz{The $b$ direction is shown here for direct comparison with the reference calculation of Ref. \citenum{zhang2023effect}; the spectra along the $a$ and $c$ axes are given in the Supporting Information.}
}
\label{fig:bse_GW}
\end{figure}

Figure~\ref{fig:bse_GW} shows the optical absorption spectra using both static and dynamical BSE calculations based on $G_0W_0$ QP energies. 
We first observe that the exciton binding energies obtained from BSE@$G_0W_0$ and the scissored BSE@DFT+$\Delta$ approach are nearly identical while the excitation energies differ considerably. 
However, the peak intensity in the BSE@$G_0W_0$ spectrum is relatively smaller compared to the BSE@DFT+$\Delta$ result. This is partly because the $G_0W_0$ quasiparticle corrections are state-specific and $k$-dependent, in contrast to the constant energy shift of the scissor operator used in DFT+$\Delta$ calculation. 
This non-uniform shift in the $G_0W_0$ calculation modifies the joint density of states, as confirmed by comparing the independent particle approximation (IPA) spectra generated from $G_0W_0$ quasiparticle energies versus scissor-shifted DFT KS eigenvalues (see Figs.  \ref{fig:bse_dft} and \ref{fig:bse_GW}).    
Notably, comparing the static and dynamical BSE@$G_0W_0$ results shows that the inclusion of dynamical screening effects leads to a redshift in the excitonic peaks by approximately 0.12 eV, from 4.39 eV to 4.27 eV. 
This redshift represents an increased exciton binding energy (Table 
\ref{table:bse_GW}), indicating a strengthened electron-hole interaction in the dynamical BSE. 

\rz{As discussed in the Theoretical Methods section, Shindo's approximation (Eq. \ref{eq:shindo}) is valid
in the limit of weak dynamical screening such that the exciton binding energy is small compared to the plasma frequency (17.9 eV). 
This effective dielectric function approach was examined against the exact diagonalization of Eq. \ref{eq:dynamic_BSE} by \citeauthor{zhang2023effect}\cite{zhang2023effect} at the level of BSE@DFT+$\Delta$ method for tuning the quasi-particle gap and with the input exciton binding energy of 1.00 eV. 
Our BSE@$G_0W_0$ method also yields a similar value for the exciton binding energy of 0.98 eV, and the dynamical contribution gives an additional 0.12 eV. 
The BSE@$G_0W_0$ calculation reported here therefore remains within this regime where the approach was validated.
Solving the two-frequency polarization function (Eq. \ref{eq:dble_freq_p}) without Shindo approximation would be necessary when the dynamical screening is significant, but it is beyond the scope of this study.}



\section{Conclusions}

In this work, we generalized the effective dielectric function method for dynamical BSE calculation, originally formulated in plane-wave (PW) basis~\cite{zhang2023effect}, to an all-electron framework based on numerical atomic orbitals (NAO) for extended periodic systems. 
This approach was demonstrated with our recently developed NAO-based BSE@$GW$ implementation~\cite{zhou2024all}. 
To reduce the substantial computational cost of dense Brillouin-zone sampling required in BSE calculations, 
we used a symmetry-adaptation strategy that maps the auxiliary basis terms from the full Brillouin zone (BZ) to the irreducible BZ (IBZ). 
We validated our NAO-based implementation by comparing dynamical BSE results obtained with scissor-shifted DFT eigenvalues against reference PW-based results using crystalline naphthalene. 
The excellent agreement, with exciton binding energies differing by less than $30$~meV, validates our NAO-based implementation. 
As a practical demonstration of the method for future application, we performed the dynamical BSE@$G_0W_0$ calculation to examine the effect of dynamical screening by comparing the static and dynamical BSE results on the crystalline naphthalene.
In future works, 
we aim to further enhance the scalability and efficiency of our all-electron periodic BSE@$GW$ framework, particularly for the dense BZ sampling, which is necessary for converging optical absorption spectra\cite{zhou2024all}. 
We will focus on incorporating advanced numerical techniques, such as iterative eigensolvers~\cite{alliati2022double,hillenbrand2025energy}, coarse-grained $k$-point grid interpolation~\cite{deslippe2012berkeleygw}, Wannier interpolation~\cite{kammerlander2012speeding}, and additional symmetry-adapted methods~\cite{stohler2026efficient}, thereby extending the applicability of BSE@$GW$ method to increasingly complex materials.


\section{Acknowledgment}

We thank the Research Computing at the University of North Carolina at Chapel Hill for computational resources. We thank Dr. Xiao Zhang and Prof. Andre Schleife  for helpful discussions.

\section{Author Contributions}
R.Z. and S.L. led and contributed equally to the work. All authors discussed the results and contributed to the manuscript writing.

\section{Data and Software Availability}

All input files supporting the findings of this study for the DFT, GW, static BSE, and dynamic BSE calculations, are provided in the Supporting Information, along with the information of the FHI-aims electronic structure package used for the development. 
The new method developed in this work (dynamical BSE) is scheduled to be integrated into the next official release of the FHI-aims software.

\bibliography{ref}

@article{blum2009ab,
  title={Ab initio molecular simulations with numeric atom-centered orbitals},
  author={Blum, Volker and Gehrke, Ralf and Hanke, Felix and Havu, Paula and Havu, Ville and Ren, Xinguo and Reuter, Karsten and Scheffler, Matthias},
  journal={Computer Physics Communications},
  volume={180},
  number={11},
  pages={2175--2196},
  year={2009},
  publisher={Elsevier}
}

@article{ren2021all,
  title={All-electron periodic {$G_0W_0$} implementation with numerical atomic orbital basis functions: Algorithm and benchmarks},
  author={Ren, Xinguo and Merz, Florian and Jiang, Hong and Yao, Yi and Rampp, Markus and Lederer, Hermann and Blum, Volker and Scheffler, Matthias},
  journal={Physical Review Materials},
  volume={5},
  number={1},
  pages={013807},
  year={2021},
  publisher={APS}
}

@article{ihrig2015accurate,
  title={Accurate localized resolution of identity approach for linear-scaling hybrid density functionals and for many-body perturbation theory},
  author={Ihrig, Arvid Conrad and Wieferink, J{\"u}rgen and Zhang, Igor Ying and Ropo, Matti and Ren, Xinguo and Rinke, Patrick and Scheffler, Matthias and Blum, Volker},
  journal={New Journal of Physics},
  volume={17},
  number={9},
  pages={093020},
  year={2015},
  publisher={IOP Publishing}
}

@article{ren2012resolution,
  title={Resolution-of-identity approach to Hartree--Fock, hybrid density functionals, RPA, MP2 and GW with numeric atom-centered orbital basis functions},
  author={Ren, Xinguo and Rinke, Patrick and Blum, Volker and Wieferink, J{\"u}rgen and Tkatchenko, Alexandre and Sanfilippo, Andrea and Reuter, Karsten and Scheffler, Matthias},
  journal={New Journal of Physics},
  volume={14},
  number={5},
  pages={053020},
  year={2012},
  publisher={IOP Publishing}
}

@article{deslippe2012berkeleygw,
  title={BerkeleyGW: A massively parallel computer package for the calculation of the quasiparticle and optical properties of materials and nanostructures},
  author={Deslippe, Jack and Samsonidze, Georgy and Strubbe, David A and Jain, Manish and Cohen, Marvin L and Louie, Steven G},
  journal={Computer Physics Communications},
  volume={183},
  number={6},
  pages={1269--1289},
  year={2012},
  publisher={Elsevier}
}

@article{rohlfing1998electron,
  title={Electron-hole excitations in semiconductors and insulators},
  author={Rohlfing, Michael and Louie, Steven G},
  journal={Physical review letters},
  volume={81},
  number={11},
  pages={2312},
  year={1998},
  publisher={APS}
}

@article{rohlfing2000electron,
  title={Electron-hole excitations and optical spectra from first principles},
  author={Rohlfing, Michael and Louie, Steven G},
  journal={Physical Review B},
  volume={62},
  number={8},
  pages={4927},
  year={2000},
  publisher={APS}
}

@article{hybertsen1986electron,
  title={Electron correlation in semiconductors and insulators: Band gaps and quasiparticle energies},
  author={Hybertsen, Mark S and Louie, Steven G},
  journal={Physical Review B},
  volume={34},
  number={8},
  pages={5390},
  year={1986},
  publisher={APS}
}

@article{hirata1999time,
  title={Time-dependent density functional theory within the Tamm--Dancoff approximation},
  author={Hirata, So and Head-Gordon, Martin},
  journal={Chemical Physics Letters},
  volume={314},
  number={3-4},
  pages={291--299},
  year={1999},
  publisher={Elsevier}
}

@article{strinati1984effects,
  title={Effects of dynamical screening on resonances at inner-shell thresholds in semiconductors},
  author={Strinati, G},
  journal={Physical Review B},
  volume={29},
  number={10},
  pages={5718},
  year={1984},
  publisher={APS}
}

@incollection{casida1995time,
  title={Time-dependent density functional response theory for molecules},
  author={Casida, Mark E},
  booktitle={Recent Advances In Density Functional Methods: (Part I)},
  pages={155--192},
  year={1995},
  publisher={World Scientific}
}

@article{loos2020dynamical,
  title={Dynamical correction to the Bethe--Salpeter equation beyond the plasmon-pole approximation},
  author={Loos, Pierre-Fran{\c{c}}ois and Blase, Xavier},
  journal={The Journal of Chemical Physics},
  volume={153},
  number={11},
  year={2020},
  publisher={AIP Publishing}
}

@article{adler1962quantum,
  title={Quantum theory of the dielectric constant in real solids},
  author={Adler, Stephen L},
  journal={Physical Review},
  volume={126},
  number={2},
  pages={413},
  year={1962},
  publisher={APS}
}

@article{wiser1963dielectric,
  title={Dielectric constant with local field effects included},
  author={Wiser, Nathan},
  journal={Physical Review},
  volume={129},
  number={1},
  pages={62},
  year={1963},
  publisher={APS}
}

@article{bloch1929quantenmechanik,
  title={{\"U}ber die {Quantenmechanik} der {Elektronen} in {Kristallgittern}},
  author={Bloch, Felix},
  journal={Zeitschrift f{\"u}r Physik},
  volume={52},
  number={7-8},
  pages={555--600},
  year={1929},
  publisher={Springer}
}

@article{albrecht1998ab,
  title={Ab initio calculation of excitonic effects in the optical spectra of semiconductors},
  author={Albrecht, Stefan and Reining, Lucia and Del Sole, Rodolfo and Onida, Giovanni},
  journal={Physical review letters},
  volume={80},
  number={20},
  pages={4510},
  year={1998},
  publisher={APS}
}

@article{kammerlander2012speeding,
  title={Speeding up the solution of the Bethe-Salpeter equation by a double-grid method and Wannier interpolation},
  author={Kammerlander, David and Botti, Silvana and Marques, Miguel AL and Marini, Andrea and Attaccalite, Claudio},
  journal={Physical Review B},
  volume={86},
  number={12},
  pages={125203},
  year={2012},
  publisher={APS}
}

@article{liu2020all,
  title={All-electron ab initio Bethe-Salpeter equation approach to neutral excitations in molecules with numeric atom-centered orbitals},
  author={Liu, Chi and Kloppenburg, Jan and Yao, Yi and Ren, Xinguo and Appel, Heiko and Kanai, Yosuke and Blum, Volker},
  journal={The Journal of Chemical Physics},
  volume={152},
  number={4},
  year={2020},
  publisher={AIP Publishing}
}

@article{blase2020bethe,
  title={The Bethe--Salpeter equation formalism: From physics to chemistry},
  author={Blase, Xavier and Duchemin, Ivan and Jacquemin, Denis and Loos, Pierre-Fran{\c{c}}ois},
  journal={The Journal of Physical Chemistry Letters},
  volume={11},
  number={17},
  pages={7371--7382},
  year={2020},
  publisher={ACS Publications}
}

@article{yao2022all,
  title={All-electron BSE@ GW method for K-edge core electron excitation energies},
  author={Yao, Yi and Golze, Dorothea and Rinke, Patrick and Blum, Volker and Kanai, Yosuke},
  journal={Journal of Chemical Theory and Computation},
  volume={18},
  number={3},
  pages={1569--1583},
  year={2022},
  publisher={ACS Publications}
}

@article{golze2019gw,
  title={The GW compendium: A practical guide to theoretical photoemission spectroscopy},
  author={Golze, Dorothea and Dvorak, Marc and Rinke, Patrick},
  journal={Frontiers in chemistry},
  volume={7},
  pages={377},
  year={2019},
  publisher={Frontiers Media SA}
}

@article{onida2002electronic,
  title={Electronic excitations: density-functional versus many-body Green’s-function approaches},
  author={Onida, Giovanni and Reining, Lucia and Rubio, Angel},
  journal={Rev. Mod. Phys.},
  volume={74},
  number={2},
  pages={601},
  year={2002},
  publisher={APS}
}

@article{marsili2017large,
  title={Large-scale G W-BSE calculations with N 3 scaling: Excitonic effects in dye-sensitized solar cells},
  author={Marsili, Margherita and Mosconi, Edoardo and De Angelis, Filippo and Umari, Paolo},
  journal={Physical Review B},
  volume={95},
  number={7},
  pages={075415},
  year={2017},
  publisher={APS}
}

@article{vorwerk2019bethe,
  title={Bethe--Salpeter equation for absorption and scattering spectroscopy: implementation in the exciting code},
  author={Vorwerk, Christian and Aurich, Benjamin and Cocchi, Caterina and Draxl, Claudia},
  journal={Electronic Structure},
  volume={1},
  number={3},
  pages={037001},
  year={2019},
  publisher={IOP Publishing}
}

@article{blase2018bethe,
  title={The Bethe--Salpeter equation in chemistry: relations with TD-DFT, applications and challenges},
  author={Blase, Xavier and Duchemin, Ivan and Jacquemin, Denis},
  journal={Chemical Society Reviews},
  volume={47},
  number={3},
  pages={1022--1043},
  year={2018},
  publisher={Royal Society of Chemistry}
}

@article{faber2014excited,
  title={Excited states properties of organic molecules: From density functional theory to the GW and Bethe--Salpeter Green's function formalisms},
  author={Faber, Carina and Boulanger, Paul and Attaccalite, Claudio and Duchemin, Ivan and Blase, Xavier},
  journal={Philosophical Transactions of the Royal Society A: Mathematical, Physical and Engineering Sciences},
  volume={372},
  number={2011},
  pages={20130271},
  year={2014},
  publisher={The Royal Society Publishing}
}

@article{jacquemin2017bethe,
  title={Is the Bethe--Salpeter formalism accurate for excitation energies? Comparisons with TD-DFT, CASPT2, and EOM-CCSD},
  author={Jacquemin, Denis and Duchemin, Ivan and Blase, Xavier},
  journal={The journal of physical chemistry letters},
  volume={8},
  number={7},
  pages={1524--1529},
  year={2017},
  publisher={ACS Publications}
}

@book{ullrich2011time,
  title={Time-dependent density-functional theory: {Concepts} and {Applications}},
  author={Ullrich, Carsten A},
  year={2011},
  publisher={OUP Oxford}
}

@article{bartlett2012coupled,
  title={Coupled-cluster theory and its equation-of-motion extensions},
  author={Bartlett, Rodney J},
  journal={Wiley Interdisciplinary Reviews: Computational Molecular Science},
  volume={2},
  number={1},
  pages={126--138},
  year={2012},
  publisher={Wiley Online Library}
}

@article{krylov2008equation,
  title={Equation-of-motion coupled-cluster methods for open-shell and electronically excited species: The hitchhiker's guide to Fock space},
  author={Krylov, Anna I},
  journal={Annu. Rev. Phys. Chem.},
  volume={59},
  pages={433--462},
  year={2008},
  publisher={Annual Reviews}
}

@article{van2013gw,
  title={The GW-method for quantum chemistry applications: Theory and implementation},
  author={van Setten, Michiel J and Weigend, Florian and Evers, Ferdinand},
  journal={Journal of chemical theory and computation},
  volume={9},
  number={1},
  pages={232--246},
  year={2013},
  publisher={ACS Publications}
}

@article{leng2016gw,
  title={GW method and Bethe--Salpeter equation for calculating electronic excitations},
  author={Leng, Xia and Jin, Fan and Wei, Min and Ma, Yuchen},
  journal={Wiley Interdisciplinary Reviews: Computational Molecular Science},
  volume={6},
  number={5},
  pages={532--550},
  year={2016},
  publisher={Wiley Online Library}
}

@article{zhou2024all,
  title={All-electron BSE@ GW method with numeric atom-centered orbitals for extended periodic systems},
  author={Zhou, Ruiyi and Yao, Yi and Blum, Volker and Ren, Xinguo and Kanai, Yosuke},
  journal={Journal of Chemical Theory and Computation},
  volume={21},
  number={1},
  pages={291--306},
  year={2025},
  publisher={ACS Publications}
}

@article{zhang2023effect,
  title={Effect of dynamical screening in the Bethe-Salpeter framework: Excitons in crystalline naphthalene},
  author={Zhang, Xiao and Leveillee, Joshua A and Schleife, Andr{\'e}},
  journal={Physical Review B},
  volume={107},
  number={23},
  pages={235205},
  year={2023},
  publisher={APS}
}

@article{cao2025applying,
  title={Applying Space-Group Symmetry to Speed Up Hybrid-Functional Calculations within the Framework of Numerical Atomic Orbitals},
  author={Cao, Yu and Zhang, Min-Ye and Lin, Peize and Chen, Mohan and Ren, Xinguo},
  journal={Journal of Chemical Theory and Computation},
  volume={21},
  number={16},
  pages={8086--8105},
  year={2025},
  publisher={ACS Publications}
}

@article{gao2016dynamical,
  title={Dynamical excitonic effects in doped two-dimensional semiconductors},
  author={Gao, Shiyuan and Liang, Yufeng and Spataru, Catalin D and Yang, Li},
  journal={Nano letters},
  volume={16},
  number={9},
  pages={5568--5573},
  year={2016},
  publisher={ACS Publications}
}

@book{bechstedt_many_2015,
	address = {Berlin, Heidelberg},
	isbn = {978-3-662-44593-8},
	url = {https://doi.org/10.1007/978-3-662-44593-8},
	title = {Many-{Body} {Approach} to {Electronic} {Excitations}: {Concepts} and {Applications}},
	publisher = {Springer},
	author = {Bechstedt, Friedhelm},
	year = {2015},
	doi = {10.1007/978-3-662-44593-8},
}

@article{shindo_effective_1970,
	title = {Effective {Electron}-{Hole} {Interaction} in {Shallow} {Excitons}},
	volume = {29},
	issn = {0031-9015},
	url = {https://journals.jps.jp/doi/abs/10.1143/JPSJ.29.287},
	doi = {10.1143/JPSJ.29.287},
	number = {2},
	urldate = {2025-12-16},
	journal = {Journal of the Physical Society of Japan},
	author = {Shindo, Koichi},
	month = aug,
	year = {1970},
	note = {Publisher: The Physical Society of Japan},
	pages = {287--296},
}

@article{bintrim2022full,
  title={Full-frequency dynamical Bethe--Salpeter equation without frequency and a study of double excitations},
  author={Bintrim, Sylvia J and Berkelbach, Timothy C},
  journal={The Journal of Chemical Physics},
  volume={156},
  number={4},
  year={2022},
  publisher={AIP Publishing}
}

@article{abbott2025roadmap,
  title={Roadmap on advancements of the FHI-aims software package},
  author={Abbott, Joseph W and Acosta, Carlos Mera and Akkoush, Alaa and Ambrosetti, Alberto and Atalla, Viktor and Bagrets, Alexej and Behler, J{\"o}rg and Berger, Daniel and Bieniek, Bj{\"o}rn and Bj{\"o}rk, Jonas and others},
  journal={Electronic Structure},
  year={2026},
  note={in press. DOI: \url{https://doi.org/10.1088/2516-1075/ae8067}},
  doi={10.1088/2516-1075/ae8067}
}

@article{grimme2006semiempirical,
  title={Semiempirical GGA-type density functional constructed with a long-range dispersion correction},
  author={Grimme, Stefan},
  journal={Journal of computational chemistry},
  volume={27},
  number={15},
  pages={1787--1799},
  year={2006},
  publisher={Wiley Online Library}
}

@article{perdew1996generalized,
  title={Generalized gradient approximation made simple},
  author={Perdew, John P and Burke, Kieron and Ernzerhof, Matthias},
  journal={Physical review letters},
  volume={77},
  number={18},
  pages={3865},
  year={1996},
  publisher={APS}
}

@article{dadsetani2015ab,
  title={Ab initio study of the optical properties of crystalline phenanthrene, including the excitonic effects},
  author={Dadsetani, Mehrdad and Nejatipour, Hajar and Ebrahimian, Ali},
  journal={Journal of Physics and Chemistry of Solids},
  volume={80},
  pages={67--77},
  year={2015},
  publisher={Elsevier}
}

@article{liu2020pyrene,
  title={Pyrene-stabilized acenes as intermolecular singlet fission candidates: importance of exciton wave-function convergence},
  author={Liu, Xingyu and Tom, Rithwik and Wang, Xiaopeng and Cook, Cameron and Schatschneider, Bohdan and Marom, Noa},
  journal={Journal of Physics: Condensed Matter},
  volume={32},
  number={18},
  pages={184001},
  year={2020},
  publisher={IOP Publishing}
}

@article{wang2016effect,
  title={Effect of crystal packing on the excitonic properties of rubrene polymorphs},
  author={Wang, Xiaopeng and Garcia, Taylor and Monaco, Stephen and Schatschneider, Bohdan and Marom, Noa},
  journal={CrystEngComm},
  volume={18},
  number={38},
  pages={7353--7362},
  year={2016},
  publisher={Royal Society of Chemistry}
}

@article{hummer2005oligoacene,
  title={Oligoacene exciton binding energies: Their dependence on molecular size},
  author={Hummer, Kerstin and Ambrosch-Draxl, Claudia},
  journal={Physical Review B—Condensed Matter and Materials Physics},
  volume={71},
  number={8},
  pages={081202},
  year={2005},
  publisher={APS}
}

@article{bhattacharya2026proton,
  title={Proton Quantum Effects on Electronic Excitation in Hydrogen-Bonded Organic Solid: A First-Principles Green’s Function Theory Study},
  author={Bhattacharya, Sampreeti and Xu, Jianhang and Zhou, Ruiyi and Kanai, Yosuke},
  journal={The Journal of Physical Chemistry Letters},
  year={2026},
  publisher={ACS Publications}
}

@article{qiu2016screening,
  title={Screening and many-body effects in two-dimensional crystals: Monolayer MoS 2},
  author={Qiu, Diana Y and Da Jornada, Felipe H and Louie, Steven G},
  journal={Physical Review B},
  volume={93},
  number={23},
  pages={235435},
  year={2016},
  publisher={APS}
}

@article{zhu2015exciton,
  title={Exciton binding energy of monolayer WS2},
  author={Zhu, Bairen and Chen, Xi and Cui, Xiaodong},
  journal={Scientific reports},
  volume={5},
  number={1},
  pages={9218},
  year={2015},
  publisher={Nature Publishing Group UK London}
}

@article{ugeda2014giant,
  title={Giant bandgap renormalization and excitonic effects in a monolayer transition metal dichalcogenide semiconductor},
  author={Ugeda, Miguel M and Bradley, Aaron J and Shi, Su-Fei and Da Jornada, Felipe H and Zhang, Yi and Qiu, Diana Y and Ruan, Wei and Mo, Sung-Kwan and Hussain, Zahid and Shen, Zhi-Xun and others},
  journal={Nature materials},
  volume={13},
  number={12},
  pages={1091--1095},
  year={2014},
  publisher={Nature Publishing Group UK London}
}

@article{bhattacharya2024bse,
  title={BSE@ GW Prediction of Charge Transfer Exciton in Molecular Complexes: Assessment of Self-Energy and Exchange-Correlation Dependence},
  author={Bhattacharya, Sampreeti and Li, Jiachen and Yang, Weitao and Kanai, Yosuke},
  journal={The Journal of Physical Chemistry A},
  volume={128},
  number={29},
  pages={6072--6083},
  year={2024},
  publisher={ACS Publications}
}

@article{tiago2005first,
  title={First-principles GW--BSE excitations in organic molecules},
  author={Tiago, Murilo L and Chelikowsky, James R},
  journal={Solid state communications},
  volume={136},
  number={6},
  pages={333--337},
  year={2005},
  publisher={Elsevier}
}

@article{holzer2021gw,
  title={The GW/BSE method in magnetic fields},
  author={Holzer, Christof and Pausch, Ansgar and Klopper, Wim},
  journal={Frontiers in Chemistry},
  volume={9},
  pages={746162},
  year={2021},
  publisher={Frontiers Media SA}
}

@article{holzer2025guide,
  title={A Guide to Molecular Properties from the Bethe--Salpeter Equation},
  author={Holzer, Christof and Franzke, Yannick J},
  journal={The Journal of Physical Chemistry Letters},
  volume={16},
  number={16},
  pages={3980--3990},
  year={2025},
  publisher={ACS Publications}
}

@article{franzke2022nmr,
  title={NMR coupling constants based on the Bethe--Salpeter equation in the GW approximation},
  author={Franzke, Yannick J and Holzer, Christof and Mack, Fabian},
  journal={Journal of Chemical Theory and Computation},
  volume={18},
  number={2},
  pages={1030--1045},
  year={2022},
  publisher={ACS Publications}
}

@article{loos2020quest,
  title={The quest for highly accurate excitation energies: A computational perspective},
  author={Loos, Pierre-Fran{\c{c}}ois and Scemama, Anthony and Jacquemin, Denis},
  journal={The journal of physical chemistry letters},
  volume={11},
  number={6},
  pages={2374--2383},
  year={2020},
  publisher={ACS Publications}
}

@article{graml2025optical,
  title={Optical excitations in nanographenes from the {Bethe--Salpeter} equation and time-dependent density functional theory: {Absorption} spectra and spatial descriptors},
  author={Graml, Maximilian and Wilhelm, Jan},
  journal={Phys. Rev. B},
  year={2026},
  volume={113},
  pages={205152},
  doi={10.1103/38k2-d55h},
  note={DOI: \url{https://doi.org/10.1103/38k2-d55h}}
}

@article{hillenbrand2025energy,
  title={Energy-specific Bethe--Salpeter equation implementation for efficient optical spectrum calculations},
  author={Hillenbrand, Christopher and Li, Jiachen and Zhu, Tianyu},
  journal={The Journal of Chemical Physics},
  volume={162},
  number={17},
  year={2025},
  publisher={AIP Publishing}
}

@article{stohler2026efficient,
  title={Efficient all-electron Bethe-Salpeter implementation using crystal symmetries},
  author={St{\"o}hler, J{\"o}rn and Bl{\"u}gel, Stefan and Friedrich, Christoph},
  journal={Phys. Rev. B},
  year={2026},
  note={in press. DOI: \url{https://doi.org/10.1103/zvb7-htlc}},
  doi={10.1103/zvb7-htlc}
}

@article{bechstedt1992efficient,
  title={An efficient method for calculating quasiparticle energies in semiconductors},
  author={Bechstedt, F and Del Sole, R and Cappellini, Giancarlo and Reining, Lucia},
  journal={Solid state communications},
  volume={84},
  number={7},
  pages={765--770},
  year={1992},
  publisher={Elsevier}
}

@article{schleife2011electronic,
  title={Electronic and optical properties of {$\mathrm{Mg}_x\mathrm{Zn}_{1-x}\mathrm{O}$} and {$\mathrm{Cd}_x\mathrm{Zn}_{1-x}\mathrm{O}$} from ab initio calculations},
  author={Schleife, Andr{\'e} and R{\"o}dl, Claudia and Furthm{\"u}ller, J{\"u}rgen and Bechstedt, Friedhelm},
  journal={New Journal of Physics},
  volume={13},
  number={8},
  pages={085012},
  year={2011}
}

@article{schleife2009optical,
  title={Optical and energy-loss spectra of MgO, ZnO, and CdO from ab initio many-body calculations},
  author={Schleife, A and R{\"o}dl, C and Fuchs, F and Furthm{\"u}ller, J and Bechstedt, F},
  journal={Physical Review B—Condensed Matter and Materials Physics},
  volume={80},
  number={3},
  pages={035112},
  year={2009},
  publisher={APS}
}

@article{alliati2022double,
  title={Double k-grid method for solving the Bethe-Salpeter equation via Lanczos approaches},
  author={Alliati, Ignacio M and Sangalli, Davide and Gr{\"u}ning, Myrta},
  journal={Frontiers in Chemistry},
  volume={9},
  pages={763946},
  year={2022},
  publisher={Frontiers Media SA}
}

@article{liu2025many,
  title={Many-body effects at heterogeneous interfaces from first-principles: Progress, challenges, and opportunities},
  author={Liu, Zhen-Fei},
  journal={ACS nano},
  volume={19},
  number={6},
  pages={5861--5870},
  year={2025},
  publisher={ACS Publications}
}

@article{wen2026dynamically,
  title={Dynamically Corrected Bethe-Salpeter Equation Solver for Self-consistent $ GW $ Reference on the Matsubara Frequency Axis},
  author={Wen, Ming and Harsha, Gaurav and Zgid, Dominika},
  journal={arXiv preprint arXiv:2604.22187},
  year={2026}
}

@article{tolle2021subsystem,
  title={Subsystem-based {GW/Bethe--Salpeter} equation},
  author={Tölle, Johannes and Deilmann, Thorsten and Rohlfing, Michael and Neugebauer, Johannes},
  journal={Journal of Chemical Theory and Computation},
  volume={17},
  number={4},
  pages={2186--2199},
  year={2021},
  publisher={ACS Publications}
}

@article{sharifzadeh2018many,
  title={Many-body perturbation theory for understanding optical excitations in organic molecules and solids},
  author={Sharifzadeh, Sahar},
  journal={Journal of Physics: Condensed Matter},
  volume={30},
  number={15},
  pages={153002},
  year={2018},
  publisher={IOP Publishing}
}

@article{hirose2015all,
  title={All-electron GW+ Bethe-Salpeter calculations on small molecules},
  author={Hirose, Daichi and Noguchi, Yoshifumi and Sugino, Osamu},
  journal={Physical Review B},
  volume={91},
  number={20},
  pages={205111},
  year={2015},
  publisher={APS}
}

@article{alvertis2023importance,
  title={Importance of nonuniform Brillouin zone sampling for ab initio Bethe-Salpeter equation calculations of exciton binding energies in crystalline solids},
  author={Alvertis, Antonios M and Champagne, Aur{\'e}lie and Del Ben, Mauro and da Jornada, Felipe H and Qiu, Diana Y and Filip, Marina R and Neaton, Jeffrey B},
  journal={Physical Review B},
  volume={108},
  number={23},
  pages={235117},
  year={2023},
  publisher={APS}
}

@article{marini2003dynamical,
  title={Dynamical excitonic effects in metals and semiconductors},
  author={Marini, Andrea and Del Sole, Rodolfo},
  journal={Physical review letters},
  volume={91},
  number={17},
  pages={176402},
  year={2003},
  publisher={APS}
}

@article{schafer1986approach,
  title={An approach to the nonequilibrium theory of highly excited semiconductors},
  author={Sch{\"a}fer, W and Treusch, J},
  journal={Z. Phys. B: Condens. Matter},
  volume={63},
  number={4},
  pages={407--426},
  year={1986},
  publisher={Springer}
}
\end{document}